\documentclass[12pt, draftclsnofoot, onecolumn]{IEEEtran}
\usepackage{graphicx,amssymb,amsmath,subcaption,amsthm}
\usepackage{multicol}
\usepackage[noadjust]{cite}
\usepackage{enumerate}
\usepackage{setspace}
\usepackage{stfloats}
\usepackage{midfloat}
\usepackage[normal]{threeparttable}
\usepackage{amsthm}
\usepackage{overpic}
\usepackage{cuted}
\usepackage{cases,subeqnarray}
\usepackage{bm,multirow,bigstrut}
\usepackage{latexsym,bm}
\usepackage{booktabs,changebar}
\usepackage{xcolor}
\usepackage{algorithmic}

\theoremstyle{plain}

\theoremstyle{plain}

\def\Htran{\mbox{\tiny $\mathrm{H}$}}
\def\Ttran{\mbox{\tiny $\mathrm{T}$}}
\def\CN{\mathcal{CN}} 
\newcommand{\vect}[1]{\mathbf{#1}}

\def\b0{{\pmb{0}}}

\def\ba{{\mathbf{a}}}

 \def\bn{{\mathbf{n}}}  
  \def\bs{{\mathbf{s}}} 
   \def\bx{{\mathbf{x}}}
\def\by{{\mathbf{y}}}

   \def\bH{{\mathbf{H}}}
\def\bI{{\mathbf{I}}}

  \def\bW{{\mathbf{W}}} 
 \def\bZ{{\mathbf{Z}}}

\newcommand{\C}{\mathbb{C}}

\IEEEoverridecommandlockouts
\begin{document}
\title{\textcolor{black}{Prospective Multiple Antenna Technologies for Beyond 5G}}
\author{Jiayi Zhang,~\IEEEmembership{Member,~IEEE}, Emil~Bj\"{o}rnson,~\IEEEmembership{Senior Member,~IEEE}, Michail Matthaiou,~\IEEEmembership{Senior Member,~IEEE}, Derrick Wing Kwan Ng,~\IEEEmembership{Senior Member,~IEEE}, Hong Yang,~\IEEEmembership{Senior Member,~IEEE}, and \\David J. Love,~\IEEEmembership{Fellow,~IEEE}
\thanks{The work of J. Zhang was supported by National Natural Science Foundation of China under Grant 61971027, Beijing Natural Science Foundation under Grants 4182049 and L171005, Science and Technology Key Project of Guangdong Province China under Grant 2019B010157001, and the ZTE Corporation. The work of E.~Bj\"{o}rnson was supported by ELLIIT and the Swedish Research Council. The work of M. Matthaiou was supported by a research grant from the Department for the Economy Northern Ireland under the US-Ireland R\&D Partnership Programme and by the EPSRC, U.K., under Grant EP/P000673/1.  The work of D. Love was supported in part by the National Science Foundation (NSF) under grant CCF1816013. The work of D. W. K. Ng was supported by funding from the UNSW Digital Grid Futures Institute, UNSW, Sydney, under a cross-disciplinary fund scheme and by the Australian Research Council's Discovery Project (DP190101363). (\textit{Corresponding Author: Jiayi Zhang})}
\thanks{J. Zhang is with the School of Electronic and Information Engineering, Beijing Jiaotong University, Beijing 100044, P. R. China. (e-mail: jiayizhang@bjtu.edu.cn).}
\thanks{E. Bj\"{o}rnson is with the Department of Electrical Engineering (ISY), Link\"{o}ping University, Link\"{o}ping, Sweden. (e-mail: emil.bjornson@liu.se).}
\thanks{M. Matthaiou is with the Institute of Electronics, Communications and Information Technology (ECIT), Queen's University Belfast, UK, BT3 9DT. (e-mail: m.matthaiou@qub.ac.uk).}
\thanks{D. W. K. Ng is with the School of Electrical Engineering and Telecommunications, University of New South Wales, NSW 2052, Australia. (e-mail: w.k.ng@unsw.edu.au).}
\thanks{H. Yang is with Nokia Bell Labs, Murray Hill, NJ 07974 USA. (e-mail: h.yang@nokia-bell-labs.com).}
\thanks{D. J. Love is with the School of Electrical and Computer Engineering, Purdue University, West Lafayette, IN 47907 USA. (e-mail: djlove@purdue.edu)}
}

\maketitle

\begin{abstract}
\textcolor{black}{Multiple antenna technologies have attracted much research interest for several decades and have
gradually made their way into mainstream communication systems. Two main benefits are adaptive
beamforming gains and spatial multiplexing, leading to high data rates per user and per cell, especially
when large antenna arrays are adopted. Since multiple antenna technology has become a key component
of the fifth-generation (5G) networks, it is time for the research community to look for new multiple antenna
technologies to meet the immensely higher data rate, reliability, and traffic demands in the beyond 5G era.
Radically new approaches are required to achieve orders-of-magnitude improvements in these metrics.
There will be large technical challenges, many of which are yet to be identified. In this paper, we survey three
new multiple antenna technologies that can play key roles in beyond 5G networks: cell-free massive
MIMO, beamspace massive MIMO, and intelligent reflecting surfaces. For each of these technologies,
we present the fundamental motivation, key characteristics, recent technical progresses, and provide
our perspectives for future research directions. The paper is not meant to be a survey/tutorial of a mature subject, but rather serve as a catalyst to encourage more research and experiments in these multiple antenna
technologies.}
\end{abstract}

\IEEEpeerreviewmaketitle

\section{Introduction}

\IEEEPARstart{T}HE demand for higher data rates and traffic volumes seems to be never-ending, thus the quest for delivering the required services must \textcolor{black}{also} continue. The cellular network technology has evolved from using fixed sector antennas to flexible multiple antenna solutions. Recently, the first release of 5G New Radio (NR) was finished by the 3rd Generation Partnership Project (3GPP) and the first commercial networks are already operational. \textcolor{black}{In particular, massive multiple-input multiple-output (MIMO), which are defined in \cite{Sanguinetti2019a} as having base stations with i) at least $64$ antennas and  ii) a number of antennas at least an order of magnitude more than the number of user equipments (UEs), is a key technology.} However, this is not the end of the MIMO development, but only the end of the beginning. As access to wireless connectivity becomes critical in our everyday lives, our expectations \textcolor{black}{of} ubiquitous coverage and service quality continue to grow.
Many future requirements that can be conceived which cannot be addressed by 5G; for example, exceptionally high bit rates, \textcolor{black}{uniform user performance over the coverage area,} ultra-low latencies, great energy efficiency, robustness against blocking and jamming, and wireless charging.

\textcolor{black}{There is no simple way to meet these requirements. There has been \textcolor{black}{significant focus on using millimeter wave (mmWave) frequencies} in 5G, since large unused bandwidths are available in these \textcolor{black}{frequency bands}, which might translate into higher bit rates.
Unfortunately, there are some fundamental drawbacks with mmWave communications \cite{jain2019impact,zhang2018low}. First, the sensitivity to signal blockage has not been resolved, despite significant research efforts \textcolor{black}{that have been devoted to the issue} in the past decade. Second, the shorter wavelength in mmWave bands \textcolor{black}{leads to} a reduced coherence time, thus one has to multiplex fewer data signals than in sub-6 GHz bands to achieve the same signaling overhead for channel state information (CSI) acquisition. For example, even if $10$ times more bandwidth can be utilized in mmWave, the bit rate might not increase if $10$ times fewer data signals can be multiplexed. These problems presumably become worse in the sub-terahertz (THz) bands, above $0.1$ THz, that are currently being studied for beyond 5G.
\textcolor{black}{ The \textcolor{black}{bottom line} is that there is a need to develop novel multiple antenna technologies that can be applied in the valuable sub-6 GHz spectrum as well as in higher bands, and to consider  both time-division duplex (TDD) and frequency-division duplex (FDD) modes.}}

It is time to analyze what lies beyond 5G, or rather what the current \textcolor{black}{multiple antenna technologies can potentially} be evolved into beyond what is currently envisaged. Potential paradigm shifts in wireless network design for beyond 5G are \textcolor{black}{cell-free massive MIMO}, beamspace massive MIMO, and intelligent reflecting surfaces (IRSs). These topics are covered in Section~\ref{section:cell-free}, Section~\ref{section:beamspace}, and Section~\ref{sec:IRS}, respectively.
New roles that these multiple antenna technologies can play for unmanned aerial vehicle (UAV)-supported communication and in sub-THz bands are discussed in Section~\ref{sec:MIMO-meets-others}, while the main conclusions are provided in Section~\ref{sec:conclusions}.
\textcolor{black}{There are several tutorial papers on multiple antenna technologies, e.g. \cite{mietzner2009multiple,Rusek2013a,rappaport2013millimeter,zheng2015survey,Xiao2017a}, and also textbooks such as \cite{Tse2005a,clerckx2013mimo,Marzetta2016a,massivemimobook,Heath2018a}. These provide an excellent introduction to the topic and also describe the \textcolor{black}{developments} that lead to 5G. When it comes to beyond-5G technologies, \cite{Renzo2019a,Bjornson2019d} are two recent papers that describe prospective future technologies, but without providing any mathematical details. In contrast, this paper provides a comprehensive description of the state-of-the-art in three selected topics and includes all the theoretical details that are essential to conduct research on these topics. Besides, various open research problems are discussed which sheds light on the development of multiple antenna technologies for beyond-5G networks. }

\section{Cell-Free Massive MIMO}
\label{section:cell-free}

The 5G cellular technology can provide data rates and traffic volumes far above previous cellular technologies, and will also reduce the latency \textcolor{black}{of} the data connections \cite{book:Kwan_5G}. Yet, these improvements are primarily achieved by \textcolor{black}{UEs} that happen to be located \textcolor{black}{near} the cell centers, while the inter-cell interference and handover issues that inherent to the cellular architecture will remain to limit the cell-edge performance. To address these issues, beyond-5G networks need to enter the cell-free paradigm, where the absence of cell boundaries alleviates the inter-cell interference and handover issues but \textcolor{black}{also gives rise to new challenges}.

\subsection{Motivation}
The first cellular networks were introduced in the 1970s to achieve more efficient use of the limited radio resources \cite{Macdonald1979a}. \textcolor{black}{Instead of requiring the data source to wirelessly communicate directly with the UE, which might be located far away and thus require very high transmit power, cellular networks consist of set of geographically distributed fixed access points (APs). The data source sends its data to a nearby AP using relatively low power. The AP forwards the data to an AP that is near the UE (typically over cables or other wireless bands) and can send the data to the UEs with relatively low power.} This enabled better spatial reuse of the frequency spectrum and the AP densification has been the main way for cellular networks to handle higher traffic volumes \cite{Cooper2010a}. However, the AP densification also leads to more inter-cell interference and more frequent handovers. Most of the traffic congestion in cellular networks nowadays is at the cell edges, since cell-center UEs can easily run their preferred applications thanks to \textcolor{black}{their lower interference levels and higher achievable} data rates. \textcolor{black}{The so-called $95$\%-likely user data rates, which can be guaranteed to $95$\% of the users and thus defines} the user-experienced performance \cite{ITU-IMT2020}, remain mediocre in 5G networks.

The solution to these issues might be to connect each user with a multitude of APs \cite{Shamai2001a,Venkatesan2007a}; if there were only one huge cell in the network, there is by definition no inter-cell interference and no need for handovers.
This solution has been explored in the past, using names such as network MIMO \cite{Venkatesan2007a,Caire2010b}, distributed MIMO \cite{Simeone2008a}, and coordinated multi-point (CoMP) \cite{Marsch2011a}. However, the practical implementation requires immense fronthaul signaling for CSI and data sharing, as well as huge computational complexity. To reduce the fronthaul signaling and computational complexity, a common approach was to divide the network into disjoint clusters containing a few neighboring APs \cite{Marsch2008a,Zhang2009b,Huang2009b}, so that only those need to exchange CSI and data. This \emph{network-centric} approach can provide some performance gains \cite{Boldi2011a}, but only partially address the interference and handover issues, which remain along the cluster edges.

The key to fully resolve these issues is to let each user be served by those APs that can reach it with non-negligible interference \cite{Bjornson2011a,Kaviani2012a,Baracca2012b}. This creates a \emph{user-centric network} \cite{Bjornson2013d}, where each AP collaborates with different sets of APs when serving different UEs. It is the UEs that select which set of APs that it is best served by, not the network. Early experiments with cell-free networks are described in \cite{Perlman2015a}, but it is first in recent years that the concept has gained \textcolor{black}{significant} traction from academia \cite{interdonato2019ubiquitous,zhang2019cell}, where the name \emph{cell-free massive MIMO} has been coined \cite{yang2013capacity,ngo2017cell,Nayebi2017a}. In a nutshell, it is a combination of the best aspects of network MIMO that were conceived in the last decade and the analytical framework from the massive MIMO literature, recently surveyed in the textbooks \cite{Marzetta2016a,massivemimobook}.

\subsection{Basics of Cell-Free Massive MIMO}

\begin{figure}
        \centering
	\begin{overpic}[width=0.7\columnwidth,tics=10]{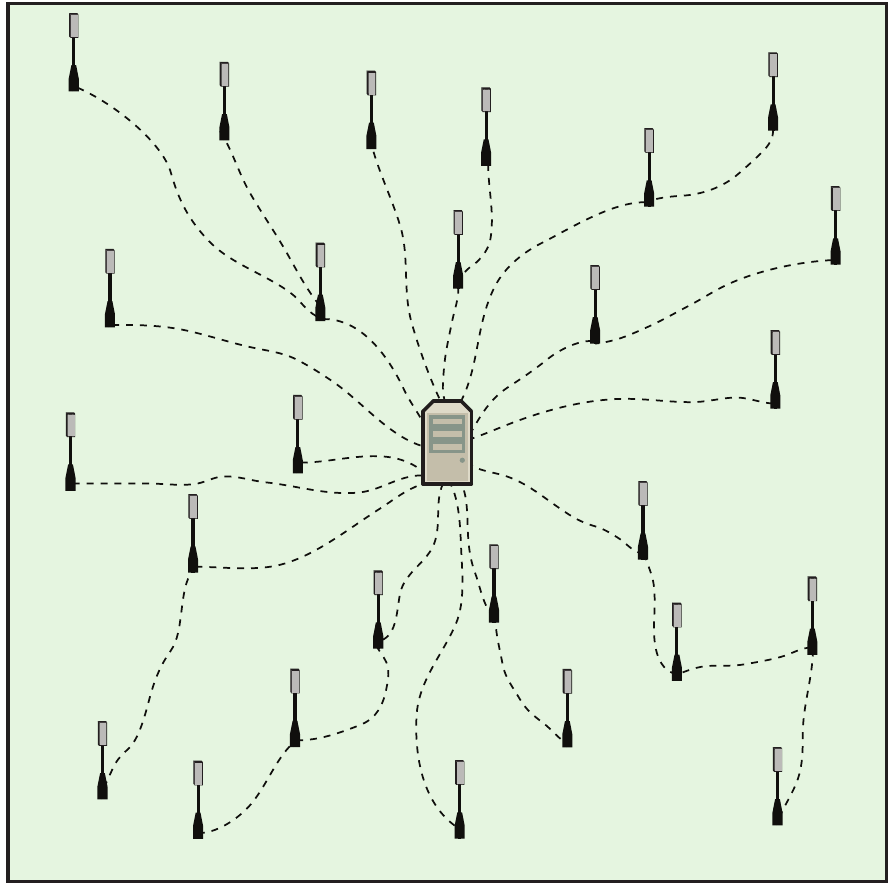}
		\put(54,92){\footnotesize{AP $l$}}
		\put(2,33){\footnotesize{Fronthaul}}
		\put(2,29){\footnotesize{connection}}
		\put(8,27.5){\vector(1,-1){7}}
		\put(54,48){\footnotesize{CPU}}
\end{overpic}
        \caption{Illustration of the network architecture in cell-free massive MIMO.}
        \label{fig:cell-free-basic}
\end{figure}
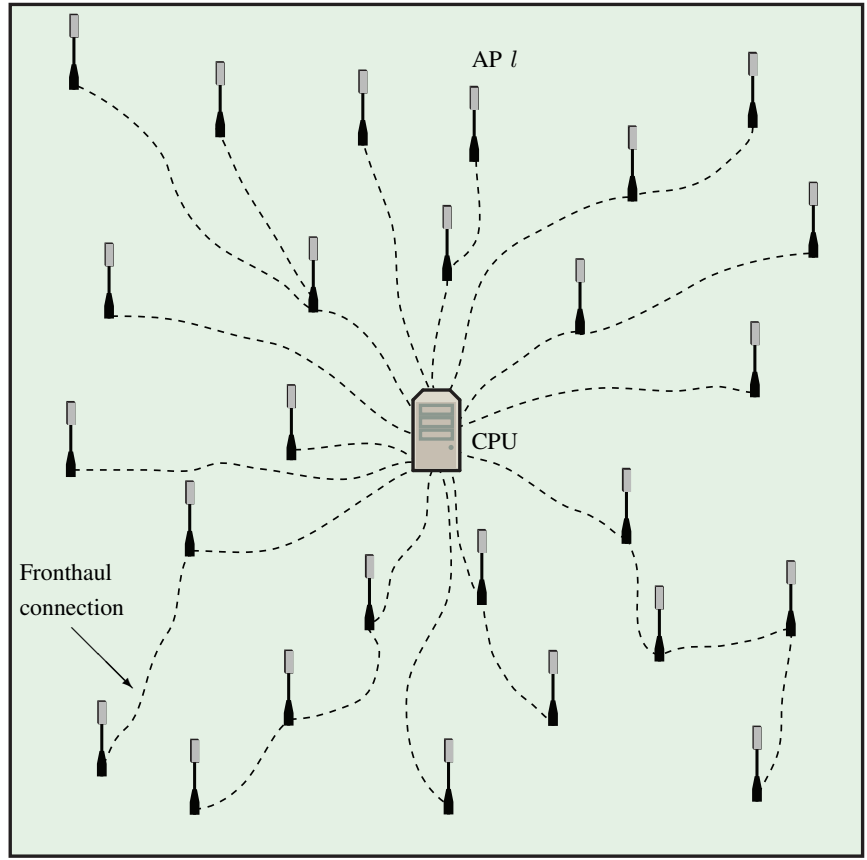

A cell-free massive MIMO network consists of a large number of APs that jointly and coherently serves a much smaller number of UEs on the same time-frequency resource. The network operates in TDD mode and exploits uplink-downlink channel reciprocity \cite{ngo2017cell,Nayebi2017a}, so that each AP can acquire CSI between itself and all UEs from uplink pilots. This CSI is sufficient to implement coherent transmission and reception \cite{Bjornson2010c}, so only data signals must be shared between APs. To enable such information flows, the APs are assumed to be connected via fronthaul to \textcolor{black}{cloud-edge} processors that take care of data encoding and decoding. These are often called central processing units (CPUs) in the literature and the structure is reminiscent of the cloud radio access network (C-RAN) architecture \textcolor{black}{\cite{Peng2016a}}. One can thus view C-RAN as an enabler of cell-free massive MIMO. The CPUs are normally assumed to only know the long-term channel qualities, while only the APs have instantaneous CSI. Fig.~\ref{fig:cell-free-basic} shows the basic network architecture of a cell-free massive MIMO system.

The spectral efficiencies that UEs can achieve in cell-free massive MIMO have been analyzed in a series of previous works. The original papers \cite{ngo2017cell,Nayebi2017a} considered single-antenna APs, single-antenna UEs, Rayleigh fading channels, and infinite capacity error-free fronthaul connections.
Later works have studied more realistic setups, such as single-antenna APs with Rician fading channels \cite{ngo2018performance,ozdogan2019performance}, multi-antenna APs with uncorrelated \cite{Chen2018b,Bashar2018a} or correlated \cite{fan2019performance,Bjornson2019c} fading, multi-antenna UEs \cite{Buzzi2017a,mai2018cell}, and hardware impairments \cite{Zhang2018a,zheng2020efficient}. The impact of finite-resolution fronthaul connections \textcolor{black}{(i.e., when both CSI and the received signal must be quantized)} was considered in \cite{Bashar2018a}. The general conclusion is that cell-free massive MIMO works well in all these cases, thus it is suitable for a variety of deployment scenarios.

One performance benchmark for cell-free massive MIMO is a cellular network with the same set of APs, but where each user is only served by one AP (i.e., a small-cell network). The first paper on the topic showed that cell-free massive MIMO can achieve nearly fivefold improvement in terms of $95$\%-likely per-user spectral efficiency \cite{ngo2017cell}.
If both the APs and the UEs are equipped with multiple antennas, the $95$\%-likely per-user performance can be further enhanced \cite{mai2018cell}. Another relevant benchmark is conventional cellular massive MIMO, consisting of a relatively small number of APs, each equipped with a large number of antennas. Such comparisons have been carried out in \cite{yang2018energy,Bjornson2019c} and show that cell-free massive MIMO can substantially improve the $95$\%-likely per-user spectral efficiency, while cellular massive MIMO is the preferred choice for cell-center UEs. This \textcolor{black}{emphasizes} the point that the cell-free paradigm is not about achieving higher peak performance, but a more uniform performance within the coverage area.
\textcolor{black}{A massive macro-diversity gain is achieved by having many geographically distributed antennas; the average distance between a UE and the closest APs reduces and the shadow fading is also combatted by the diversity.
Moreover,} the energy efficiency of cell-free massive MIMO was considered in \cite{ngo2017total,yang2018energy}, which showed that it can improve the energy efficiency (measured in bit/Joule) by nearly ten times compared to cellular massive MIMO.
Hence, two main reasons to adopt cell-free massive MIMO in beyond-5G networks is the vastly higher $95$\%-likely spectral efficiency and energy efficiency.

\subsection{System Model and Key Characteristics}

\begin{figure}[t!]
	\centering
	\begin{overpic}[width=1\columnwidth,tics=10]{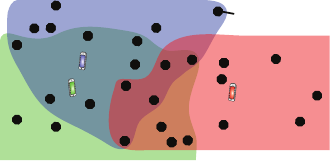}
	\put(72,43){AP $l$}
	\put(27,29){User 1}
	\put(20,45){AP cluster for user 1}
	\put(9.5,22){User 2}
	\put(1,1){AP cluster for user 2}
	\put(73,19){User 3}
	\put(64,4){AP cluster for user 3}
\end{overpic}
	\caption{Example of how different subsets  $\mathcal{M}_k$ of APs serve three UEs ($k=1,2,3$) in a cell-free massive MIMO system.}
	\label{fig:illustrateCooperation}
\end{figure}
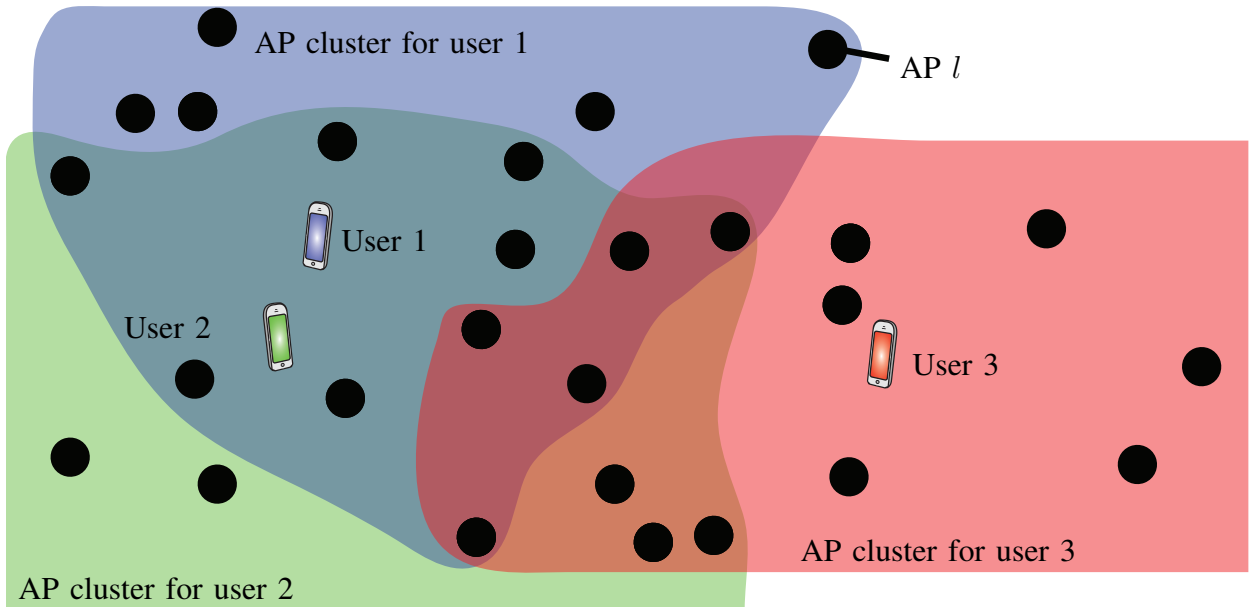

We will now explain the key characteristics of cell-free massive MIMO in further detail by considering a basic system model. We assume there are $L$ APs in the network, each equipped with $N$ antennas, and $K$ single-antenna UEs. User $k$ is served by a subset $\mathcal{M}_k \subset \{1,\ldots,L\}$ of the APs, which have been selected in a user-centric manner. Fig.~\ref{fig:illustrateCooperation} exemplifies how the APs can be divided into partially overlapping subsets when serving the UEs.

The channel between AP $l$ and user $k$ is denoted by $\vect{h}_{kl} \in \mathbb{C}^{N}$, and it is the same uplink and downlink due to the TDD operation. The UEs send uplink pilots that enable AP $l$ to compute local estimates $\hat{\vect{h}}_{kl}$ of the channels to all UEs ($k=1,\ldots,K$) \cite{liu2020tabu}. Different channel estimators can be used depending on the channel model, but we will not cover those details to keep the description general and short; we refer the interested readers to \cite{mai2018cell,Bjornson2019Aug,ozdogan2019performance}. Deep learning can also be used to estimate the channel \cite{jin2019channel,jin2020channel}. \textcolor{black}{Irrespective of the choice of estimator, pilot contamination appears in cell-free massive MIMO (just as in any large-scale network). Fortunately, it appears to be less of a concern than in cellular massive MIMO since each AP has few antennas and only serves a few UEs.}

\subsubsection{Uplink Data Transmission}

Let \textcolor{black}{$x_{k}$} denote the unit-power signal that user $k$ wants to \textcolor{black}{send over} the uplink. The user assigns a transmit power $p_k \geq 0$ to the signal and transmits it simultaneously with all other UEs, thereby expecting that the network can spatially separate the UEs' signals.
The received uplink signal ${\bf y}_{l}^{\rm{ul}} \in \mathbb{C}^N$ at AP $l$ becomes
\setcounter{equation}{0}
\begin{align} \label{eq:uplink-signal}
{\bf y}_{l}^{\rm{ul}} = \sum_{i=1}^{K} \vect{h}_{il} \sqrt{p_{i}} x_i + \vect{n}_k,
\end{align}
where $\vect{n}_k \sim \CN(\vect{0},\sigma^2 \vect{I}_N)$ is the complex-valued independent additive white Gaussian noise
and $\bI_{N}$ denotes the $N\times N$ identity matrix.
Note that \eqref{eq:uplink-signal} is a summation of the UEs' signals received over different channels.
The APs with indices in $\mathcal{M}_k$ will use their received signals $\{ {\bf y}_{l}^{\rm{ul}} : l \in \mathcal{M}_k\}$ to jointly detect the signal transmitted from user $k$. More precisely, each AP selects a receive combining vector $\vect{v}_{kl} \in \mathbb{C}^{N}$ and computes the inner product $\vect{v}_{kl}^{\Htran} {\bf y}_{l}^{\rm{ul}}$, where $(\cdot)^{\Htran}$ denotes the conjugate transpose. This scalar is then sent to the CPU which combines the contributions from all the APs that serve user $k$:
\begin{equation} \label{eq:combined-signal-UL}
\sum_{l \in \mathcal{M}_k} \vect{v}_{kl}^{\Htran} {\bf y}_{l}^{\rm{ul}} = \sum_{l \in \mathcal{M}_k}\sum_{i=1}^{K}  \vect{v}_{kl}^{\Htran}  \vect{h}_{il} \sqrt{p_{i}} x_i +  \sum_{l \in \mathcal{M}_k} \vect{v}_{kl}^{\Htran} \vect{n}_k .
\end{equation}
By following the standard methodology from the massive MIMO literature \cite[Theorem~4.4]{massivemimobook}
to compute a lower bound on the uplink capacity, an achievable spectral efficiency for user $k$ is
\begin{align} \label{eq:uatf-ul}
\mathrm{SE}_{k}^{({\rm ul})} =  \log_2  \left( 1 + \mathrm{SINR}_{k}^{({\rm ul})}  \right),
\end{align}
where $\mathrm{SINR}_{k}^{({\rm ul})}$  can be interpreted as an effective signal-to-interference-and-noise ratio (SINR) and is given by

\begin{align}
\mathrm{SINR}_{k}^{({\rm ul})}&=\frac{ p_{k} \left|  \mathbb{E} \left\{ \sum\limits_{l \in \mathcal{M}_k} \vect{v}_{kl}^{\Htran} \vect{h}_{kl} \right\} \right|^2  }{
\sum\limits_{i=1}^{K} p_{i}  \mathbb{E} \left\{\left|  \sum\limits_{l \in \mathcal{M}_k} \vect{v}_{kl}^{\Htran} \vect{h}_{il}\right|^2 \right\}
-p_{k} \left|  \mathbb{E} \left\{ \sum\limits_{l \in \mathcal{M}_k} \vect{v}_{kl}^{\Htran} \vect{h}_{kl} \right\} \right|^2 + \sigma^2 \sum\limits_{l \in \mathcal{M}_k} \mathbb{E}\{ \|  \vect{v}_{kl}  \|^2\}} \label{eq:uplink-instant-SINR-level2}.
\end{align}

The combining vectors are to be selected at the respective APs based on locally available CSI, which means that AP $l$ should select $\vect{v}_{kl}$ as a function of the estimates $\{ \hat{\vect{h}}_{il}: i =1,\ldots,K \}$ from itself to the different UEs. Before describing some common combining methods, we divide the vector into two parts:
\setcounter{equation}{4}
\begin{equation} \label{eq:cell-free-weights}
\vect{v}_{kl} = a_{kl} \frac{\bar{\vect{v}}_{kl}}{\sqrt{\mathbb{E} \{ \| \bar{\vect{v}}_{kl} \|^2\}}},
\end{equation}
where $a_{kl} \in \mathbb{C}$ is a deterministic weighting factor and \textcolor{black}{$\bar{\vect{v}}_{kl} / \sqrt{\mathbb{E} \{ \|\bar{\vect{v}}_{kl} \|^2\}}$} is a unit-power combining vector that depends on the CSI.
The purpose of the weighting factors is that APs with good channel conditions should get higher weights and thereby have a large influence on the combined signal in \eqref{eq:combined-signal-UL}. The use of such weights is also known as large-scale fading decoding \cite{Nayebi2016a}, particularly when the weights are selected at the CPU based on channel statistics from the entire network. A general expression for the optimal weights is found in \cite{Bjornson2019c}.

The receive combining vectors can be computed in different ways.
The original paper \cite{ngo2017cell} on cell-free massive MIMO considered maximal ratio (MR) combining, where $\bar{\vect{v}}_{kl} = \hat{\vect{h}}_{kl}$. This method maximizes the received signal power without taking the existence of other UEs into account. One key benefit of using this method is that expectations in  \eqref{eq:downlink-SINR-level1} can often be computed in closed form; for example, under uncorrelated \cite{ngo2017cell} or correlated \cite{Bjornson2019c} Rayleigh fading and for Rician fading \cite{ozdogan2019performance}.
However, higher spectral efficiencies are achieved by \textcolor{black}{local minimum
mean-square error (L-MMSE) combining, for which the combining vector can be expressed as}\footnote{\color{black}In some cases, the inverse matrix in \eqref{eq:L-MMSE} also includes the covariance matrices of the channel estimation errors; see \cite{Bjornson2019c}.}
\begin{equation} \label{eq:L-MMSE}
\bar{\vect{v}}_{kl} = \bigg(\sum\limits_{i =1}^{K} p_{i} \hat{\vect{h}}_{il} \hat{\vect{h}}_{il}^{\Htran} + \sigma^2 \vect{I}_N \bigg)^{-1}\hat{\vect{h}}_{kl}.
\end{equation}
Interestingly, L-MMSE outperforms MR even in the case of single-antenna APs \cite{BjornsonPIMRC2019,Bjornson2019c}, but is not an optimal combining method since that would require all the APs to jointly selected their receive combining.

\begin{figure}
        \centering
	\begin{overpic}[width=1\columnwidth,tics=10]{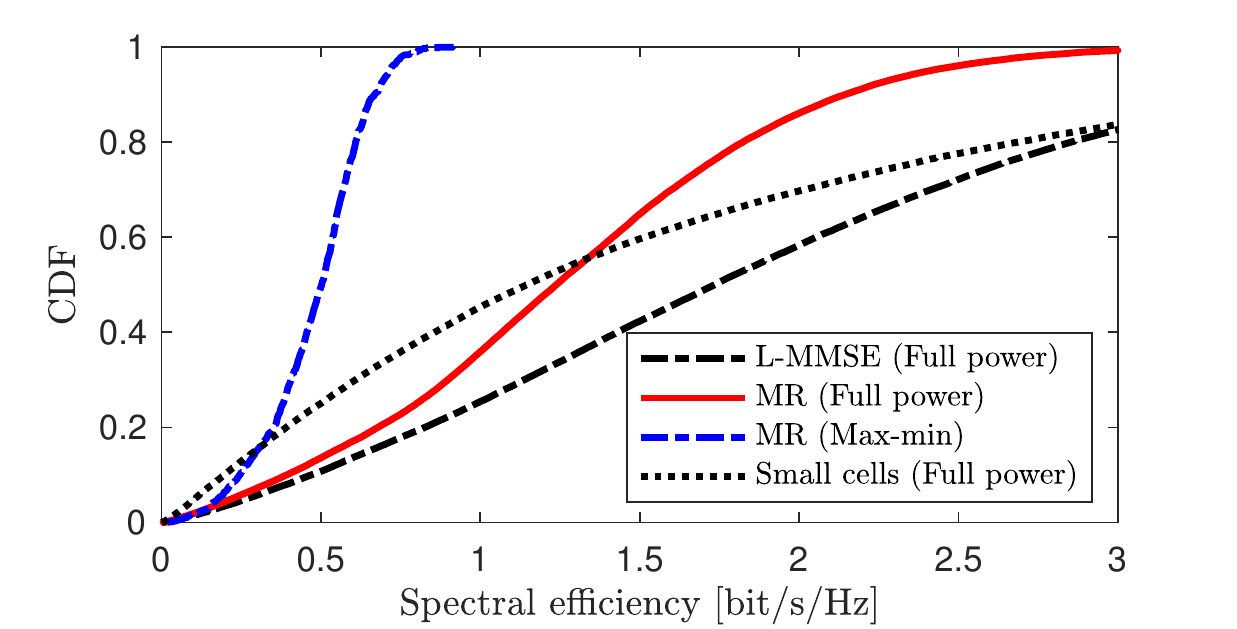}
\end{overpic}
        \caption{The uplink spectral efficiency with different combining methods and power control schemes, using the simulation setup from \cite{ngo2017cell}.}
        \label{fig:simulation_uplink_distributed}
\end{figure}

We will now illustrate the performance behaviors. Fig.~\ref{fig:simulation_uplink_distributed} shows the cumulative distribution function (CDF) of the per-user spectral efficiency in a setup with $L=100$ single-antenna APs and $K=40$ UEs uniformly distributed in a $1\times 1$ km square. We refer to \cite{ngo2017cell} for further details on the simulation parameters. The CDF is computed by considering different random realizations of the AP and user locations.

There are three curves where all UEs transmit at full power and one curve where power control is used to maximize the worst-user spectral efficiency in the network, using an algorithm from \cite{ngo2017cell}. The highest spectral efficiency is achieved by cell-free massive MIMO when using L-MMSE combining and the optimal weights from \cite{Bjornson2019c}. All UEs benefit from using that method compared to the case of small cells, where each user connects to the AP providing the highest value. It is particularly the weakest ``cell-edge'' UEs that benefit from the cell-free approach, while UEs that happen to be very close to an AP \textcolor{black}{do not} benefit much. If L-MMSE is replaced with MR, the strongest UEs (which are interference-limited) lose in performance while the weakest UEs (which are noise-limited) are barely affected. If one further applies the power-control scheme from  \cite{ngo2017cell}, the $1$\% weakest UEs get an improvement in spectral efficiency but the gain is barely visible since these UEs are noise-limited from the beginning.

The conclusion is that cell-free massive MIMO can greatly improve the performance for the weakest UEs in a network. The gains are achieved by coherent processing of the signals received at multiple APs. Power control policies can be used to shape the CDF curves in different ways. The previous works \cite{ngo2017cell,Nayebi2016a,Bashar2019a} have considered the design of power control that maximizes the worst-user performance, which might not be the desired optimization criterion since a minor performance improvement comes at the price of reducing most UEs' performance \textcolor{black}{by} a lot (as shown in Fig.~\ref{fig:simulation_uplink_distributed}). Other power control schemes need to be developed in the future. One recent example is \cite{Nikbakht2019a}.

\subsubsection{Downlink Data Transmission}

Let \textcolor{black}{$\check{x}_k$} denote the unit-power downlink signal intended for user $k$.
Each AP $l \in \mathcal{M}_k$ that serves this user maps the scalar signal to its $N$ antennas using a precoding vector $\vect{w}_{kl} \in \mathbb{C}^{N}$, thus making $\vect{w}_{kl} \check{x}_k$ the transmitted signal.
When all APs follow that procedure, the received downlink signal $y_k^{\rm{dl}} \in \mathbb{C}$ at user $k$ becomes
\begin{align}
y_k^{\rm{dl}} = \sum_{i=1}^{K} \sum_{l \in \mathcal{M}_i} \vect{h}_{kl}^{\Ttran} \vect{w}_{il} \check{x}_i + n_k,
\end{align}
where $n_k \sim \CN(0,\sigma^2)$ is independent additive noise.
By following a similar methodology as in the uplink, an achievable spectral efficiency for user $k$ is  \cite[Theorem~4.6]{massivemimobook}
\begin{equation} \label{eq:downlink-rate-expression-general}
\mathrm{SE}_{k}^{({\rm dl})} =  \log_2  \left( 1 + \mathrm{SINR}_{k}^{({\rm dl})}  \right)  \textrm{bit/s/Hz},
\end{equation}
where $\mathrm{SINR}_{k}^{({\rm dl})}$ can be interpreted as an effective SINR and is given by
\begin{small}
\begin{align}
\mathrm{SINR}_{k}^{({\rm dl})} &= \frac{ \left|   \mathbb{E} \left\{ \sum\limits_{l \in \mathcal{M}_k} \vect{h}_{kl}^{\Ttran} \vect{w}_{kl} \right\} \right|^2 }{
\sum\limits_{i=1}^{K}
  \mathbb{E} \left\{  \left| \sum\limits_{l \in \mathcal{M}_i} \vect{h}_{kl}^{\Ttran} \vect{w}_{il} \right|^2 \right\}  -   \left|   \mathbb{E} \left\{ \sum\limits_{l \in \mathcal{M}_k} \vect{h}_{kl}^{\Ttran} \vect{w}_{kl} \right\} \right|^2+ \sigma^2 }. \label{eq:downlink-SINR-level1}
\end{align}
\end{small}
The expectations  in \eqref{eq:downlink-SINR-level1} are taken with respect to random channel fading realizations. One can compute the SINR using Monte-Carlo simulations for any channel distribution and any way of selecting the precoding vectors.

As described earlier, it is preferable if AP $l$ selects its precoding vector based only on its locally available CSI, which consists of the estimates $\{ \hat{\vect{h}}_{il} : i =1,\ldots,K \}$ from itself to the different UEs. Before describing some common precoding methods, we divide the vector into two parts:
\begin{equation}
\vect{w}_{kl} = \sqrt{\rho_{kl} } \frac{\bar{\vect{w}}_{kl}}{\sqrt{\mathbb{E} \{ \| \bar{\vect{w}}_{kl} \|^2\}}},
\end{equation}
where $\rho_{kl}\geq0$ represents the transmit power that AP $l$ assigns to user $k$ and
\textcolor{black}{$\bar{\vect{w}}_{kl} / \sqrt{\mathbb{E} \{ \|\bar{\vect{w}}_{kl} \|^2\}}$} is a unit-power precoding vector that determines the spatial directivity of the signal.

Similar to the uplink, the original papers \cite{ngo2017cell,Nayebi2017a} considered MR precoding, where $\bar{\vect{w}}_{kl} = \hat{\vect{h}}_{kl}^*$, which is also known as conjugate beamforming since $(\cdot)^*$ denotes complex conjugation. This method directs the transmission towards the intended receiver without taking the existence of other UEs into account. One key benefit of using this method is that expectations in  \eqref{eq:downlink-SINR-level1} can often be computed in closed form for many common channel models \cite{Bjornson2019c,ozdogan2019performance}.
However, it has recently been shown in \cite{BjornsonPIMRC2019} that signal-to-leakage-and-noise ratio (SLNR) precoding with
\begin{equation}
\bar{\vect{w}}_{kl} = \bigg(\sum\limits_{i =1}^{K} \rho_{il} \hat{\vect{h}}_{il}^* \hat{\vect{h}}_{il}^{\Ttran} + \sigma^2 \vect{I}_N \bigg)^{-1}\hat{\vect{h}}_{kl}^*
\end{equation}
provides higher spectral efficiencies by balancing between maximizing the signal power at the intended receiver and minimizing the interference that leaks to non-intended receivers.
Interestingly, SLNR outperforms MR even in the case of single-antenna APs \cite{BjornsonPIMRC2019} and the computational complexity is almost the same.
Nevertheless, SLNR precoding is not the optimal method. Other papers consider similar methods such as full-pilot zero-forcing \cite{Interdonato2018a} and L-MMSE \cite{Bjornson2019Aug} precoding.

For brevity, we will not provide any simulation results for the downlink, because the performance of the different precoding methods are reminiscent of their uplink counterparts. However, we stress that the selection of the transmit power $\rho_{kl}$ is more complicated in the downlink than in the uplink, because each AP needs to select how to distribute their power between different UEs. Some different power allocation schemes are found in \cite{ngo2017cell,Nayebi2017a,Interdonato2019a}, but further work is still required to deal with other criteria than the maximization of the worst UEs' performance.

\subsection{Scalable Large-Scale Deployment}\label{se:scalable}

The main challenge in designing cell-free massive MIMO is to achieve a network architecture that is scalable in the sense of being implementable in a large network, spanning an entire city. One way to define scalability is to consider the limit $K \to \infty$ and evaluate if the network could still operate in that case \cite{Bjornson2019Aug}. More precisely, the computational complexity at each AP and its fronthaul capacity requirement must be independent of $K$. All the combining and precoding methods described above can be modified to satisfy that condition if each AP is only allowed to serve a fixed number of UEs, irrespective of how large the network is. This is a natural restriction since the UEs are typically distributed over the network and thereby close to different APs. It was shown in \cite{Bjornson2019Aug} that this restriction has a negligible impact on the spectral efficiency.

Other issues are harder to implement in a scalable way. Downlink power allocation is one such issue, where each AP needs to assign power to the UEs that it serves without having complete knowledge about the channel conditions that other APs are facing. Any global power allocation optimization must have a complexity that grows with $K$, thus making the implementation infeasible in large networks. As a compromise approach, various heuristic power allocation schemes can be found in \cite{Bjornson2010c,Bjornson2011a,Nayebi2017a,Interdonato2019a}, but it lies in the nature of these schemes that it is hard to evaluate how well they perform in practice. Hence, further work is necessary to understand how to perform effective and scalable power allocation. On the other hand, an uplink counterpart to downlink power allocation is the selection of the combining weights $a_{kl}$, which can be selected optimally \cite{Nayebi2016a,Bjornson2019c} but the complexity will grow with $K$. There is currently no good understanding how to select these weights in a distributed but effective manner. Uplink power control can potentially also have an important impact.

There is also a series of papers that consider partially centralized precoding and combining for cell-free massive MIMO \cite{Nayebi2016a,Bjornson2019c,Bjornson2019Aug,Yang2019a}. These methods allow for interference suppression between APs, which can substantially increase the spectral efficiency. First steps toward a scalable implementation of these methods are taken in \cite{Bjornson2019Aug}.

Another aspect is related to cloud-RAN and the fronthaul infrastructure because a large network will require multiple CPUs and the encoding/decoding tasks that need to be carried out at the CPU level must be distributed between them. Some first works in these directions are found in \cite{Perlman2015a,Burr2018a,Interdonato2019a}, but there is no quantitative comparison of different network infrastructures.

\subsection{Open Research Problems}

The above brief survey of cell-free massive MIMO provides a quick overview from its inception to the state-of-the-art. Although much research has been carried out leading to much scientific progress in recent years, there remain several important and interesting open research problems which we believe are necessarily relevant in propelling the technology to the next stage. We shall discuss three such problems, \textcolor{black}{in no} particular order, in the following.

\subsubsection{Power Control}

As already mentioned, power control, especially downlink power control, still requires much research. The optimal power control coefficients, such as the ones for max-min power control \cite{ngo2017cell,Nayebi2017a}, can be obtained using second-order cone programming. Unfortunately, these methods are not fast enough for real-time implementation. Unlike cellular systems, where downlink and uplink power controls are symmetric in the sense that each has the same number of power control coefficients, there is a different number of coefficients in cell-free networks.
There are $K$ coefficients in the uplink, while if each of the $M$ APs is serving all $K$ UEs, there are $MK$ power control coefficients in the downlink. Even when taking into account that only a subset of APs \textcolor{black}{$\mathcal{M}_k$ serves each user $k$}, there will be \textcolor{black}{$\sum_{k=1}^{K} |\mathcal{M}_k| \geq K$} coefficients to select in the downlink, where \textcolor{black}{$|\mathcal{M}_k|$} denotes the number of elements in $\mathcal{M}_k$. Hence, the computational complexity for downlink power control is particularly high.
 In a cellular system where each user is served by a single AP, practical power control works to maintain a target SINR iteratively for each user, i.e., when the target SINR for a particular user is exceeded, the allocated power for that user is lowered and vice versa. It has been shown such algorithm converges \cite{Zander1992a,foschini1993simple,yates1995framework,hanly1995algorithm,hanly1996capacity} under some very general assumptions. Nevertheless, this approach cannot be directly applied in cell-free massive MIMO where each user is simultaneously served by many APs, which must coordinate their decisions (some may increase their power while some may decrease) and must also satisfy per-AP power constraints. Furthermore, the impact of real-world power control with finite discrete power levels must also be investigated. In addition, how to use power control to effectively strike a balance between sum spectral efficiency and user fairness is also not well understood, since the max-min fairness approach may \textcolor{black}{sacrifice too much sum} spectral efficiency to provide absolute fairness to all UEs. While the proportional fairness metric is often used in cellular networks \textcolor{black}{\cite{huaizhou2013fairness,massivemimobook}}, the picture can be quite different in  cell-free networks, where the UEs' SINRs are distributed in a very different way. When it comes to the uplink, there is evidence \cite{Yang2019a,Bjornson2019c} that full power transmission might work well in many practical scenarios. This marks a significant departure from cellular thinking where full power transmission does not work for the uplink \cite{Yang2014a}. We should mention that although there are only $K$ power control coefficients, they can interact with the \textcolor{black}{$\sum_{k=1}^{K} |\mathcal{M}_k|$} weights in \eqref{eq:cell-free-weights}. If these can be jointly optimized in a computationally efficient way, further performance improvement may be achieved.

\subsubsection{Fronthaul/Backhaul Provisioning}

With a large number of \textcolor{black}{APs} scattered across the intended coverage area, it is obvious that the burden of fronthaul/backhaul for cell-free networks will be much heavier than that in traditional cellular systems. Wired provision via optical fiber is cost-prohibitive except in premium venues or if serial connections can be used \cite{interdonato2019ubiquitous}. Wireless provision is a viable option but comes with its own set of challenges such as the availability of spectrum and all the difficulties of reliably delivering ultra-high data rates wirelessly. One idea is to use a ``dual layer'' architecture where a cellular \textcolor{black}{massive MIMO network} hauls a cell-free massive MIMO system \cite{marzetta2016dual}. For cell-free networks to become a successful reality, more out-of-the-box ideas for fronthaul/backhaul provisions are undoubtedly needed and the best C-RAN methods must be utilized. Another angle of attacking this same problem is to research the means of minimizing the fronthaul/backhaul requirements, by decentralizing the processing as far as possible and heavily quantize the signals sent over the fronthaul \cite{Bashar2018a} or using distributed quantization techniques \cite{varshney_distributed,blum_dist1,chen_tong_varshney,brown_quantizedDD,choi_quantizedDD,choi_codedDD}.
A recent overview of the trade-offs between fronthaul requirements and the uplink performance for some well-known processing schemes is provided in \cite{Bjornson2019c}. A main conclusion is that there are opportunities to exploit the specificities of the fronthaul architecture to optimize its utilization and create semi-distributed methods.

\subsubsection{Network Scalability}

One definition of ``network scalability'' from \cite{Bjornson2019Aug} was provided earlier, based on letting $K \to \infty$, but other definitions are also possible. Generally speaking, it refers to the ability to meaningfully increase the performance of a small system by an ``arbitrarily'' larger system. For massive MIMO, a larger system means a larger number of service antennas.
Taking the fronthaul/backhaul capacity and network computing power limitations into account, a skeleton for scalable cell-free massive MIMO operation was provided in \cite{Interdonato2019a,Bjornson2019Aug}.
\textcolor{black}{But,} how to optimally and seamlessly associate each user to a group of APs and select the necessary signal processing and power control \textcolor{black}{remain very challenging problems}---any attempt to globally optimize these operations would be fundamentally unscalable. Another facet of scalability is how the performance \textcolor{black}{scales with respect to an} increase of the number of APs, $L$, while keeping the number of UEs $K$ fixed. For example, it is shown in \cite{Yang2019a} that the uplink minimum data rate does not scale with $L$ when using MR processing. Scaling laws should also be investigated for real-world channels, such as with mixed \textcolor{black}{line-of-sight (LoS)} and non-LoS propagation, and other non-stationary channels \cite{ali2019linear,de2019non}. For LoS channels, approaches used in \cite{yang2017LoS,yang2018LoS} for cellular networks can be adapted for cell-free networks \cite{yang2019LoS}.

\section{Beamspace Massive MIMO}
\label{section:beamspace}

The more antennas that are used in a MIMO transceiver, and the higher the carrier frequency and bandwidth are, the more complicated the implementation becomes. \textcolor{black}{One} way to reduce the implementation complexity, without sacrificing too much in performance or operational flexibility, is to utilize the spatial structure of the channels and transceiver hardware. In this section, we describe beamspace massive MIMO, which is the general concept that underpins hybrid beamforming and its future successors. We particularly focus on recent progress and open problems related to using lens arrays for beamspace massive MIMO.

\subsection{Motivation}
Early research in single-user MIMO focused on open-loop techniques that achieve MIMO benefits without  any transmit-side knowledge of CSI.  The most popular open-loop techniques are diversity techniques (e.g. space-time block codes and space-time trellis codes) and multiplexing techniques (e.g. spatial multiplexing)
\cite{bib:Alamouti98,bib:Tarokh98}. The performance of these techniques is severely limited, in terms of most measures of performance, because the transmission is not adapted to the current CSI in any way.

In the early 2000s, the single-user MIMO research shifted to techniques that adapt the transmit signal to the channel conditions using some level of transmit-side CSI \cite{scaglione_stoica_barbarossa_giannakis_sampath,visotsky_madhow,jongren_skoglund_infofeedback,zhou_giannakis,love_heath_SM}.  \textcolor{black}{The simplest kind of this} adaptation is linear precoding, where a multiple antenna open-loop signal is adapted to the channel by multiplying it by a precoding matrix before transmission \cite{scaglione_stoica_barbarossa_giannakis_sampath}.  The benefit is that spatial CSI adaptation is now encapsulated in the precoding matrix. Linear precoding has had a widespread impact with inclusion in multiple standards including 4G \textcolor{black}{Long-Term Evolution (LTE)}, 5G NR, and  versions of  IEEE 802.11. 

Because commercially available MIMO transmitters have typically had fewer than eight antennas,  linear precoding has historically been implemented  using direct digital implementation.   For example, LTE Release 8 systems limit the base station to having at most four antenna ports.  With these small array sizes, direct digital processing of precoding was generally practical because it is cost-effective for small arrays to use a relatively high-resolution \textcolor{black}{analog-to-digital converter (ADC)} at each transmit element.  In this kind of implementation, it is convenient to think of the linear precoder as a single matrix \textcolor{black}{and the transmit signal is multiplied by that matrix}, avoiding many sophisticated formulations of precoder design.

Commercial array sizes, however, are poised to scale dramatically over the coming years.
mmWave and massive MIMO transmitters equipped with on the order of a hundred antennas are commonly discussed  in the literature.  As the number of antenna elements increases, there are multiple  benefits to rethinking the signal processing and implementation of linear precoding.  The most popular approach is using a beamspace MIMO formulation.

\textcolor{black}{We firmly believe that beamspace terminology, notation, and thinking will be critical for 5G and beyond systems. In sub-6 GHz, the number of antennas will continue to increase.  The dimensionality of these arrays combined with the unique hardware characteristics (hybrid digital-analog, sub-arraying, tiled arrays, etc.) will make it impractical, if not impossible, to sound each array element.  For this reason, MIMO processing will be best done using a subspace approach based on virtual or effective channels. At mmWave frequencies and higher, beamspace will be indispensable. These arrays will be large and \textcolor{black}{may have} non-traditional array hardware implementations.  Optical-like thinking will begin to be more important, which aligns perfectly with beamspace.}

\subsection{Signal Model}
Considering a single-user MIMO system, the standard input-output expression is
\begin{equation}\label{eq_MIMOIO}
\by  = \bH\bs + \bn,
\end{equation}
where $\by\in\C^{N_r}$ is the received signal, $\bH\in\C^{N_r\times N_t}$ is the channel matrix, \textcolor{black}{$\bs\in\C^{N_t}$} is the transmitted signal, and $\bn\in\C^{N_r}$ is additive noise (which could include multiuser interference when the model is extended to a multiple user system).  In a precoding formulation with $M$ parallel data streams, the transmit signal $\bs$ is decomposed into a linear precoder $\bW\in\C^{N_t\times M}$ and \textcolor{black}{a data} signal $\bx\in\C^{M}$ according to
\begin{equation}\label{eq_precIO}
\bs = \bW\bx.
\end{equation}
As mentioned earlier, in a fully digital implementation, the multiplication in (\ref{eq_precIO}) is performed using a standard digital processor.

A beamspace approach decomposes the precoder further as a product of two different precoders as
\begin{equation}\label{eq_precIO1}
\bW = \bW_1\bW_2,
\end{equation}
where $\bW_1 \in \C^{N_t\times N_{v,t}}$ and $\bW_2\in \C^{N_{v,t}\times M}$ for some positive integer $N_{v,t}.$
Plugging this into (\ref{eq_MIMOIO}) yields
\begin{equation}\label{eq_MIMOIO1}
\by  = \bH\bW_1\bW_2\bx + \bn.
\end{equation}
The two precoders may be constrained and selected using drastically different criteria.

\begin{figure}[t!]
	\centering
	\begin{overpic}[width=1\columnwidth,tics=10]{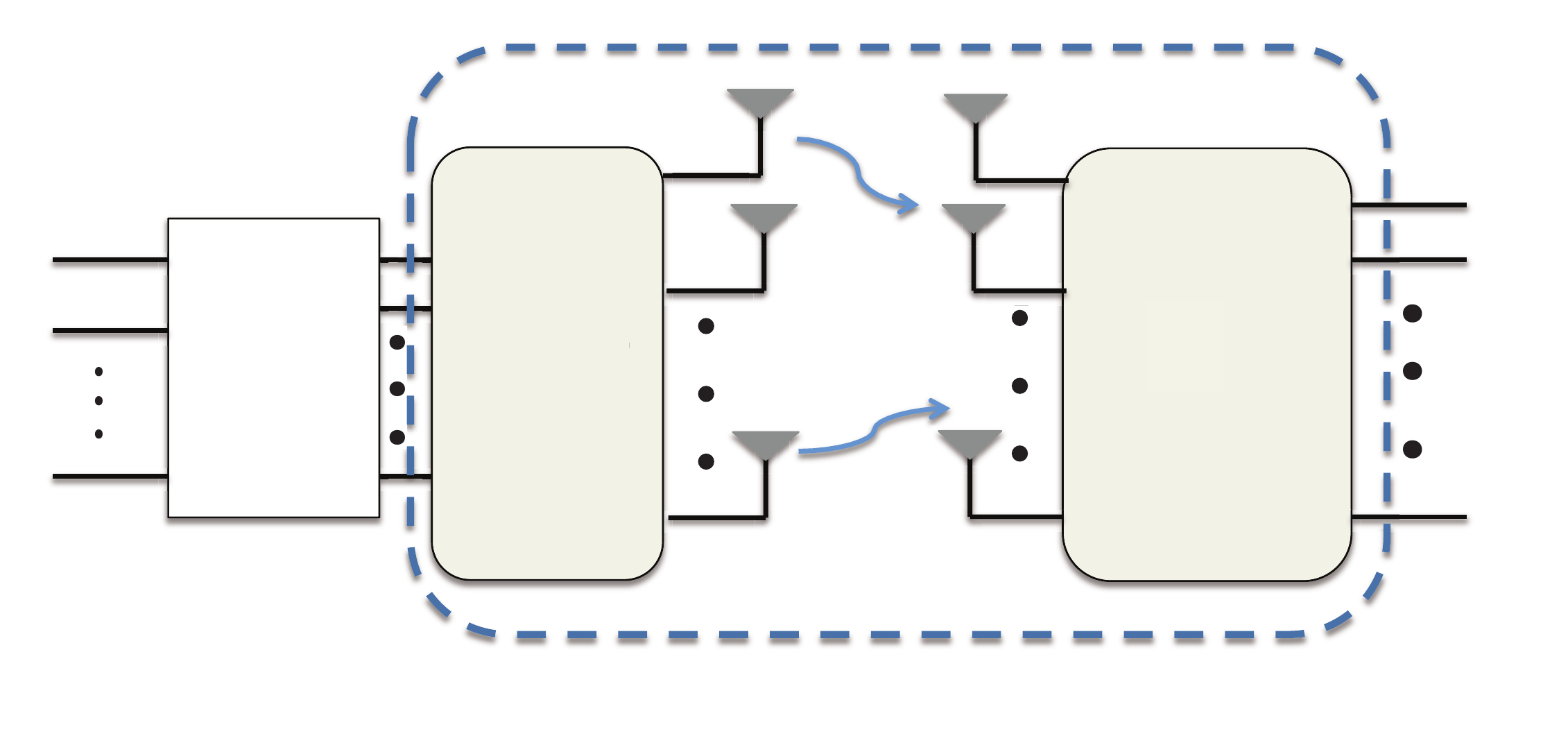}
	\put(13.5,22){$\bW_2$}
	\put(31.5,22){$\bW_1$}
	\put(74.5,22){$\bZ^{\Htran}$}
	\put(55,1){$\bH_{v}$}
	\put(53,24){$\bH$}
	\put(-1,20){$\bx$}
	\put(95,20){$\by_v$}
\end{overpic}
	\caption{Illustration of the beamspace system model in \eqref{eq:beamspace-input-output}.}
	\label{fig_beamspace}
\end{figure}
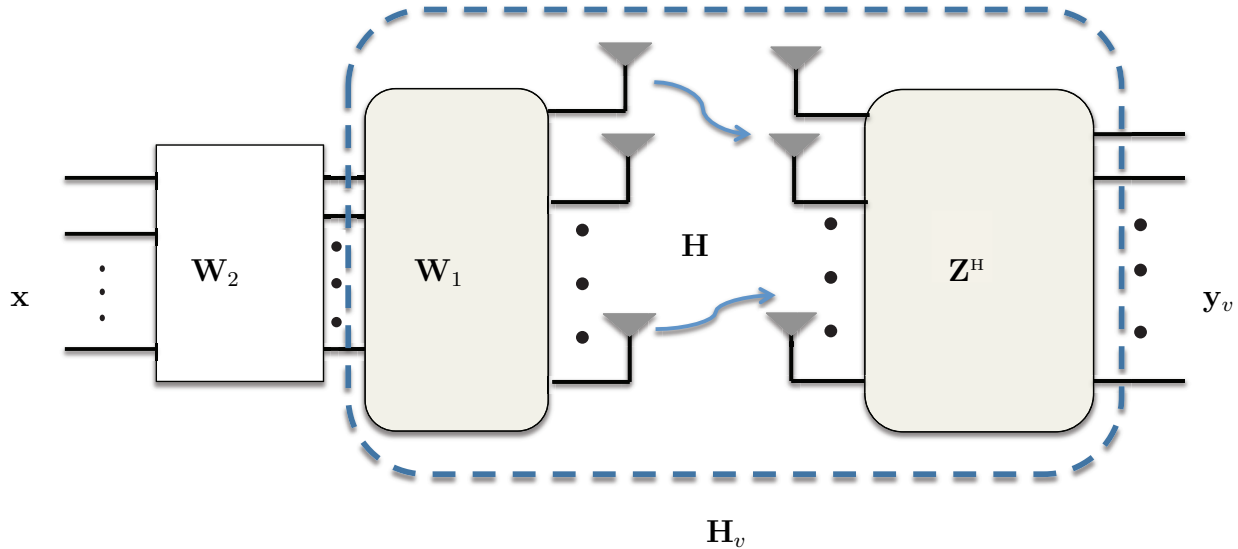

The receiver might also use linear processing.  In this setup, the receiver applies a linear combining filter $\bZ\in\C^{N_{r}\times N_{v,r}}$ and processes the received signal $\by_v =  \bZ^{\Htran}\by$. Typically, the precoder $\bW_1$ and receiver $\bZ$ are selected first.  Then, the second transmit precoder $\bW_2$ is  selected conditioned on the selected $\bW_1$ and $\bZ.$    Thus, $\bW_2$ is selected using a \textit{virtual} or \textit{effective}  (e.g. see the discussion in \cite{sayeed_deconstructing,love_heath_lau_gesbert_etal,love_heath_SM,hong_sayeed_SM})
$N_r\times N_v$ MIMO model\textcolor{black}{
\begin{equation} \label{eq:beamspace-input-output}
\by_v = \bH_v \bW_2\bx + \bn_v,
\end{equation}}where the virtual/effective channel is $\bH_{v} = \bZ^{\Htran}\bH\bW_1$ \textcolor{black}{and the virtual/effective noise is $\bn_{v} = \bZ^{\Htran}\bn$}. Note that in the case when linear receive processing is not explicitly used, the linear receiver can be implicitly assumed to be $\bZ = \bI_{N_r}$. The input-output model \eqref{eq:beamspace-input-output} is illustrated in Fig.~\ref{fig_beamspace}.

For some channel models and linear processing architectures that \textcolor{black}{exploit} the channel structure, the virtual channel $\bH_v$ can exhibit a variety of advantageous properties. A common assumption is that the virtual channel $\bH_v$ is \textit{sparse} \cite{sayeed_raghavan_sparse}, meaning that $\bH_v$ contains a limited number of non-zero entries. One example is when $\bZ$ and $\bW_1$ \textcolor{black}{respectively} contain the left and right singular vectors of $\bH$. In a practical deployment, entries will usually not be exactly zero, but the matrix can have a limited number of entries that are much larger in magnitude than the remaining entries.
This sparsity can simplify  the operation and intuitive behavior of $\bW_2$ in many situations.  In extreme cases where $\bH_v$ is diagonal, $\bW_2$ can be thought of as performing virtual subchannel selection and power loading.

\subsection{History}
Beamspace approaches have recently received much interest for application in massive MIMO and mmWave communications.  The idea of beamspace processing, however, has a long history dating back to early radar systems \cite{chapman_beamspace,richards_radar}.  Radar systems have regularly employed arrays with hundreds of elements dating back to at least the 1960s \cite{skolnik_radar}.
In cellular systems, the beamspace is widely utilized in \textcolor{black}{LTE-Advanced (LTE-A)} through the idea of dual-codebook precoding.  Release $10$ of LTE-A first included a dual codebook approach for eight-antenna downlink precoding.  The matrix $\bW_1$, often called the wideband matrix, was selected to adapt to spatial characteristics of the channel  \cite{dualcodebook_TDOC}.    The matrix $\bW_2$ was then chosen conditioned on $\bW_1$.

The idea of virtualized channel processing became central under the concept of \textit{transparency} in LTE-A. \textcolor{black}{CoMP} systems allow a UE to receive a signal sent from multiple geographically distributed transmission points, which could be utilizing diverse forms of precoding and \textcolor{black}{multiuser} transmission. To simplify the control, knowledge, and computational burden on the UE, the standard allows the UE to be configured with multiple reference signals and \textcolor{black}{CSI feedback reports}. In beamspace formulation, the UE could be configured with $K$ reference signals and \textcolor{black}{CSI feedback reports}. The multiple transmission points could send reference signals over each of the possible first precoders $\bW_1[1],\ldots,\bW_1[K].$  The sounded precoder $\bW_1[k]$ would have a corresponding virtual channel $\bH_v[k]$.  The user would then send feedback for selection of the precoder (i.e., through the corresponding \textcolor{black}{CSI feedback reports}) for each of the respective virtual channels.

This virtual approach allows operators and manufacturers to deploy sophisticated precoding schemes and easily upgrade to new precoding schemes because the user is not required to have any knowledge of $\bW_1[1],\ldots,\bW_1[K]$. The user is only required to know the number of reference signals, \textcolor{black}{CSI feedback reports}, and corresponding configuration information for each.  This kind of future-proof thinking has carried on in 3GPP for a variety of purposes. More recently, the practical use of beamspace has been reinvigorated because of the interest in hybrid beamforming and precoding at mmWave frequencies \cite{elayach_hybrid}.

\subsection{Implementation} 
The vast majority of beamspace techniques are based on phase-shifter architectures. In this approach, the first precoder is of the form
\begin{equation}
\bW_1 = \alpha_1 \left[
\begin{array}{ccc}
e^{j\phi_{0,0} } & \cdots  & e^{j\phi_{0,{N_{v,t}-1}}}   \\
\vdots  & \ddots  & \vdots  \\
  e^{j\phi_{N_t-1,0}}& \cdots   &   e^{j\phi_{N_t-1,{N_{v,t}-1}}}
\end{array}
\right],
\end{equation}
where $\alpha_1$ is a gain factor and $\{\phi_{m,n}\}$ are phases.
The approach could similarly be applied to the combining matrix $\bZ$  at the receiver.

The most common phase choices are based on the discrete Fourier transform (DFT) matrix.  In this scenario,  $\phi_{m,n} = 2\pi mn/N_t$ for $n=0,\ldots,N_t-1$ and $m=0,\ldots,N_{v,t}-1$.  These phase selections offer many benefits when the array is a uniform linear array (ULA) and far-field communication is considered.
As shown in \cite{sayeed_deconstructing}  for the case of ULAs at the transmit and receiver,  the channel $\bH$ can be  written as
\begin{equation}
\bH = \int_{-1/2}^{1/2} \int_{-1/2}^{1/2}  G(\theta_r,\theta_t) \ba_r(\theta_r)\ba_t^{\Htran}(\theta_t) d\theta_r d\theta_t,
\end{equation}
where $\theta_t$ is the normalized angle-of-departure (AoD),  $\theta_r$ is the normalized angle-of-arrival (AoD), $G(\theta_r,\theta_t) $ is the scattering function at AoD $\theta_t$ and AoA $\theta_r$, and
\begin{alignat}{2}
\ba_t(\theta_t) &= \frac{1}{\sqrt{N_t}}\left[1~e^{-j2\pi\theta_t}~\cdots ~ e^{-j2\pi(N_t-1)\theta_t} \right]^{\Ttran},\\
\ba_r(\theta_r) &= \frac{1}{\sqrt{N_r}}\left[1~e^{-j2\pi\theta_r}~\cdots ~ e^{-j2\pi(N_r-1)\theta_r} \right]^{\Ttran}.
\end{alignat}
The choice of a DFT-based precoder and linear receiver  then correspond to uniformly sampling the AoD and \textcolor{black}{AoA} spaces, respectively.

These phase shifters could be implemented digitally, but there are a variety of practical benefits that come from an analog or radio-frequency (RF) domain implementation \cite{zhang_molisch_kung,Sudarshan_metha_etal}.  In the RF domain approach, the phase shifters are applied after the digital-to-analog converter (DAC) when used at the transmitter and before the \textcolor{black}{ADC} at the receiver.  The lack of amplitude variation in the entries of $\bW_1$ means that the phase shifters can be efficiently implemented using RF integrated circuit and microwave monolith integrated circuit techniques.

In the case of DFT precoding, the precoder can be also implemented using a Butler matrix \cite{butler_matrix,molisch_zhang}.  The Butler matrix  operates for the square matrix scenario (i.e.,  $N_t = N_{v,t}$ or $N_r = N_{v,r}$).  It is implemented as a passive network using phase couplers and phase shifters.

\subsection{Beamspace Using Lens Arrays}

Recent advances in RF technology have moved away from using discrete antenna elements, making antenna arrays that function more like an optical system.
This can be achieved using \textit{lens arrays}. Among various definitions, \cite{7416205} defines a lens array as a device whose main function is to ``provide variable phase shifting for \textcolor{black}{electromagnetic (EM)} rays at different points on the lens aperture so as to achieve angle-dependent energy focusing property.'' \textcolor{black}{The lens arrays do not rely on lossy and expensive \textcolor{black}{phase shifters} and can offer nearly orthogonal beams as they act as DFT matrices. The advantages of the lens arrays over conventional phase-shifter based systems are presented in \cite{Chae_lens2}.} \textcolor{black}{Moreover, compared to lens arrays with phase shifters (see for e.g. \cite{Venugopal} and references therein), lens arrays offer substantial hardware and power savings.}

With the development of mmWave communications over the past decade, lens-based topologies have come \textcolor{black}{to} the forefront of wireless communications research \cite{JR:Lens_Atenna}. The reason is simple: By harnessing the focusing capabilities of lens arrays, \textcolor{black}{one can focus the EM} power arriving from different directions on different lens ports, thereby transforming the spatial MIMO channel into its sparse beamspace representation. Most importantly, by doing so \textcolor{black}{the system} can select only a small number of dominant beams ($\ll N_{v,t}N_{v,r}$) that carry most of the \textcolor{black}{EM energy, which reduces} the effective dimension of the MIMO channel matrix for signal processing manipulations along with the associated number of RF chains.

The first approach that combined the properties of lens arrays with the beamspace methodology can be found in \cite{5707050}. This paper proposed the concept of
the continuous aperture phased MIMO (CAP-MIMO) \textcolor{black}{architecture}, which uses a discrete lens array (DLA) to enable a quasi-continuous aperture phased MIMO operation at mmWave frequencies. The same research group published \textcolor{black}{several} papers on this topic underpinned by physical demonstrations \cite{brady_CAPMIMO,Sayeed2016}. In the following, we will overview the most recent advances in the area of lens-array based MIMO topologies and also identify some open problems that require further research.

\subsubsection{Channel estimation}
Conventional hybrid mmWave systems with high-resolution phase shifters offer much greater flexibility in the analog precoder design than \textcolor{black}{lens arrays} (e.g. using compressed sensing techniques as in \cite{elayach_hybrid,zhang_RIP}), which translates into enhanced channel estimation accuracy. Lens-based topologies are inherently inflexible in this sense since the analog precoders have to be DFT matrices. This makes the conventional channel estimation schemes tailored towards hybrid architectures with phase shifters problematic. We can categorize the channel estimation schemes for lens-based topologies that have been developed over the past years into two categories:

\begin{itemize}

\item \textit{Narrowband channel estimation:} The estimation of the narrowband beamspace MIMO channels with lens arrays was originally studied in
\cite{Hogan_Sayeed, Zhang_lens2018}. \textcolor{black}{The techniques in both those references}, though seemingly different, harness the sparsity of the beamspace channel to select only the dominant beams which capture most of the \textcolor{black}{EM} power. By doing so, the dimension of the beamspace channel is substantially reduced and this facilitates the signal processing manipulations; for instance, in \cite{Zhang_lens2018} the conventional linear \textcolor{black}{minimum mean-squared error (MMSE)} estimator was used. The weakness with the approach in \cite{Hogan_Sayeed} is that the number of pilot symbols to scan across all the beams is proportional to the number of antennas. In the massive MIMO regime, this number will scale \textcolor{black}{poorly} leaving limited resources for data transmission.
An alternative route for improving the channel estimation accuracy is through the support detection (SD)-based scheme of \cite{Dai_reliable,han_compress}.
Here, the main idea is to decompose the total channel estimation problem into a series of sub-problems each containing a sparse channel component.
As a next step, for each of these components, their support is first detected \textcolor{black}{then removed sequentially}.

\item \textit{Wideband channel estimation:} In a massive antenna array, it is very likely that the propagation
delay across the array is comparable to the symbol period. In such a case, different antenna elements will receive different time-domain symbols emanating from the same physical path at the same sampling time. This phenomenon is known in the literature as the \textit{spatial-wideband
effect} \cite{Jin_Gao2018,Jin_Gao_Rappaport2018}. With wideband signaling, such an effect will cause \textit {beam squint} in the frequency domain
meaning that the AoAs (AoDs) will become frequency-dependent\footnote{To articulate the importance of \textcolor{black}{beam squint,} one can think of an OFDM system where each subcarrier will experience distinct AoAs for the same physical path, thereby making channel estimation and transceiver design complicated exercises.}. Despite the importance of this phenomenon, to the best of our knowledge, the only relevant paper is \cite{Dai_Sayeed}, which proposed a successive support detection (SSD) technique; the main idea here is that each sparse path component has frequency-dependent support determined by its spatial direction which can be estimated using beamspace windows. Then, the authors apply the principle of serial interference cancelation on each single path component. It is also worth mentioning two earlier works in the area of wideband channel estimation for hybrid systems with phase shifters, \textcolor{black}{namely} \cite{Dai_selective} and \cite{Venugopal}. In the former, \textcolor{black}{a distributed grid matching pursuit} algorithm was proposed while in the \textcolor{black}{latter} utilized the orthogonal matching pursuit technique. Yet, \textcolor{black}{neither} of these papers considers the beam squint effect.

\end{itemize}

\textit{Open challenges:} From the above discussion, it is obvious that the area of channel estimation for lens-based topologies at mmWave frequencies is still in its infancy. We will now outline some open problems that require further investigation:
\begin{itemize}

\item [a)] Following a stream of recent papers \cite{Jin_Gao2018,Jin_Gao_Rappaport2018,Matthaiou_JSTSP}, one can recast the channel estimation problem as a channel reconstruction problem by harnessing the \textit{AoA-delay reciprocity} between the uplink and downlink in an FDD system. Hence, one only needs to regularly estimate the frequency-dependent path gains. A comprehensive performance analysis is currently missing.

\item [b)] The area of channel estimation for 3D lenses is also very timely given the importance of such geometries at higher frequencies (e.g. mmWave, sub-THz bands). A recent article on this topic is \cite{Qi_Lens2017}, which showed that the dominant entries of the channel matrix of 3D lens arrays form a dual crossing shape and then introduced an iterative algorithm that leverages this property.

\item [c)] As was previously mentioned, lens arrays offer substantial hardware and power savings compared to phase shifters. Nevertheless, the total implementation cost and power consumption of a mmWave transceiver can be further reduced by deploying coarse ADC quantizers. In such a case, the problem of channel estimation becomes far more complicated, particularly for wideband systems where different antennas at each sampling time collect non-identical data symbols \cite{Jin_Gao_Rappaport2018}. To the best of our knowledge, the only relevant paper in this space is \cite{Vlachos}, which addressed channel estimation using the Expectation-Maximization algorithm.

\end{itemize}

\subsubsection{Hardware imperfections}
Lens arrays are lossy devices, as was identified already in some early papers in the field of microwave engineering and antenna theory (e.g. \cite{McGrath}). A neat classification of the different types of losses in constrained lens arrays can be found in \cite{Popovic}. Yet, in the field of communications engineering, hardware imperfections of lens arrays is a vastly unexplored problem. We will now provide an overview of some recent contributions in this context.

\begin{figure}[t!]
\centering
\includegraphics[width=0.88\columnwidth]{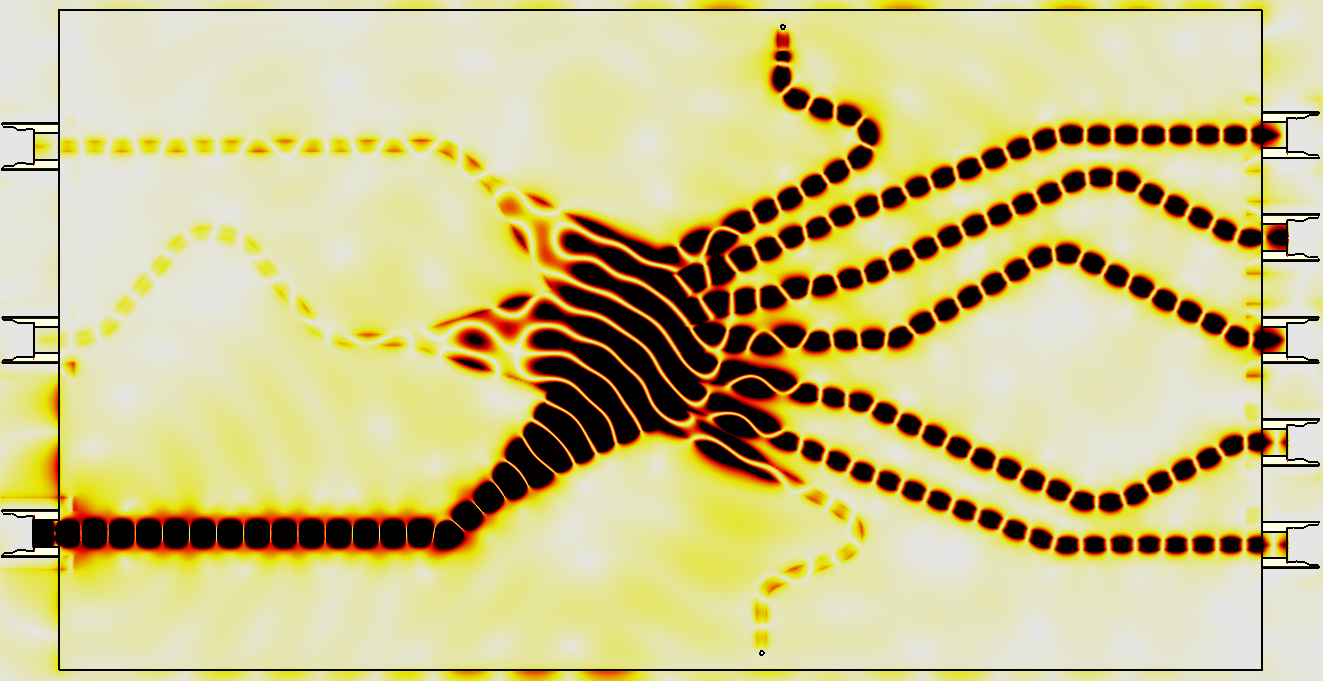}%
\caption{Electric field distribution $200$ $\mu$m inside the substrate layer of the Rotman lens in \cite{Fusco1}.}
\label{Efield}
\end{figure}

In \cite{Tataria1}, the aggregate impact of \textit{switching errors} and \textit{spillover losses} was characterized for an uplink multiuser MIMO mmWave system with a lens array at the BS. The former losses are a result of imperfect absorption and isolation characteristics of concurrent RF switches, which result in impedance mismatches and poor port-to-port isolation \cite{Fusco1}. On the other hand, spillover losses are due to the fact that the finite number of antenna elements renders the sampling of the AoAs imperfect. As such, the RF power desired for a particular beam port also leaks into neighboring beam ports. \textcolor{black}{In Fig.~\ref{Efield}, we elucidate this phenomenon by illustrating the {electric field} distribution inside the substrate of the Rotman lens\footnote{\textcolor{black}{A Rotman lens, originally proposed by Rotman and Turner in \cite{Rotman_original}, is a type of  microwave beamforming network that allows multiple antenna beams to be formed without the need for switches or phase shifters. In principle, a Rotman lens can steer the direction of the output array transmission based on  the input direction of the incoming beam, such that gains of 10 to 15 dB are obtained. Over the years, it has been successfully integrated in low-cost communications, remote-piloted vehicles, radar and satellite systems \cite{Rotman_overview}.}} of \cite{Fusco1}. The figure demonstrates the importance of spillover losses in lens arrays since a significant portion of the incoming energy is dissipated towards one of the dummy ports and the remaining portion is bounced back to other beam ports. \textcolor{black}{The numerical simulations} showed that out of 1 Watt power entering the beam port, only 0.55 Watt is calculated to leave all the array ports.}

\begin{figure}
  \centering
\begin{tabular}[c]{cc}
\hspace{-80pt}
\begin{subfigure}[b]{0.5\textwidth}
\centering
  \includegraphics[width=0.35\textwidth]{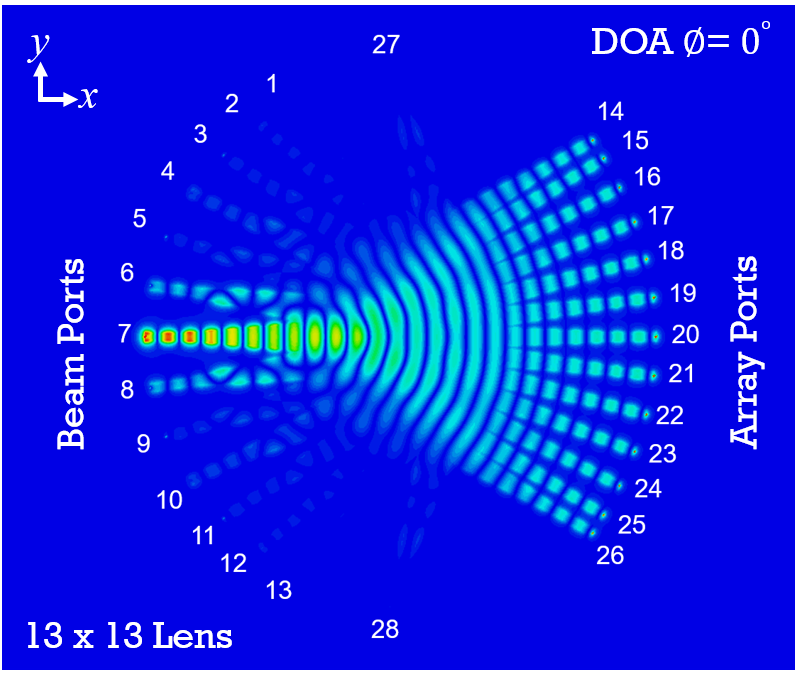}
  \caption{$\phi=0^{\circ}$}
  \label{fig:sub-first}
\end{subfigure}&\hspace{-130pt}
\begin{subfigure}[b]{0.5\textwidth}
  \centering
  \includegraphics[width=0.35\textwidth]{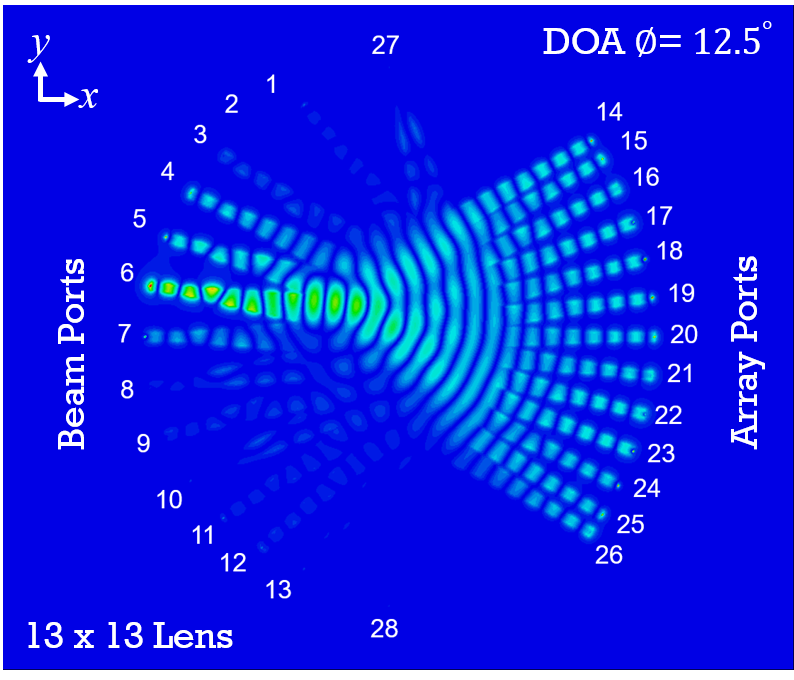}
  \caption{$\phi=12.5^{\circ}$}
  \label{fig:sub-second}
\end{subfigure}\\
\hspace{-80pt}
\begin{subfigure}[b]{0.5\textwidth}
  \centering
  \includegraphics[width=0.35\textwidth]{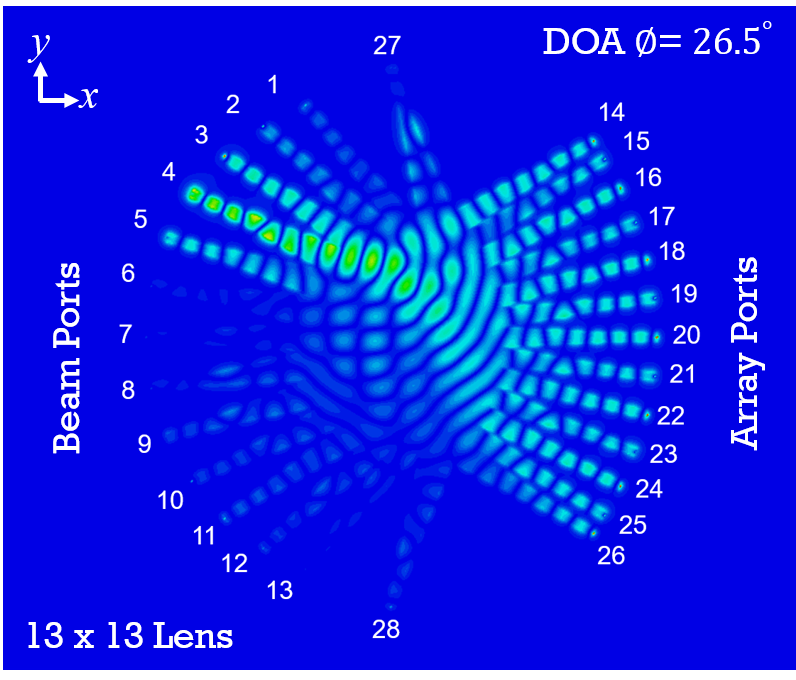}
  \caption{$\phi=26.5^{\circ}$}
  \label{fig:sub-third}
\end{subfigure}&\hspace{-130pt}
\begin{subfigure}[b]{0.5\textwidth}
  \centering
  \includegraphics[width=0.35\textwidth]{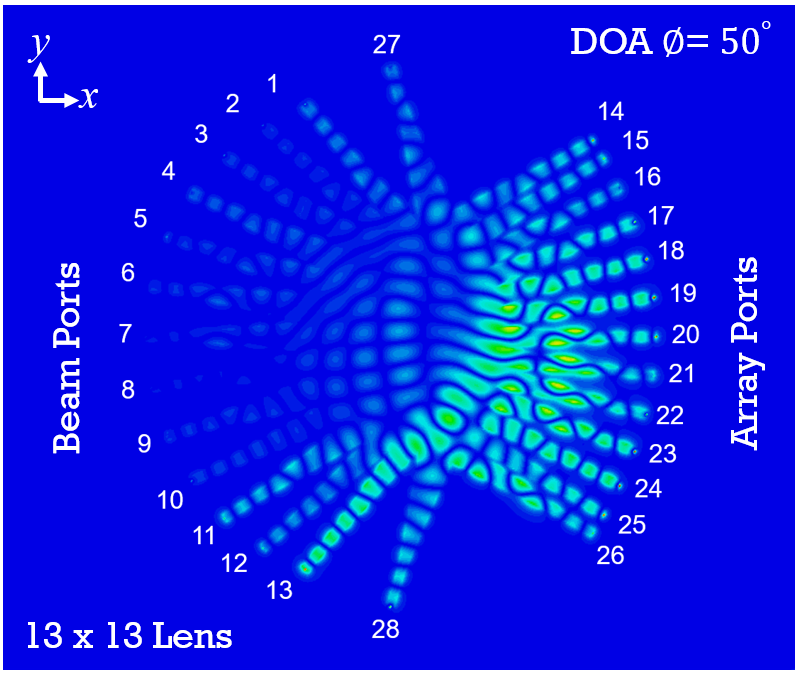}
  \caption{$\phi=50^{\circ}$}
  \label{fig:sub-fourth}
\end{subfigure}
\end{tabular}
\caption{Surface \textcolor{black}{electric field} distribution $200$ $\mu m$ inside the substrate layer of the $13\times 13$ Rotman lens at multiple AoAs, denoted by $\phi$. Data taken from \cite{Fusco2}.}
\label{fig:fig}
\end{figure}

On a similar note, \cite{Fusco2} provided a full \textcolor{black}{EM} characterization of spillover losses at $28$\,GHz and demonstrated that the \textcolor{black}{EM focusing inside a lens is more accurate towards the broadside excitation angles; see Fig.~\ref{fig:fig}, which shows the surface \textcolor{black}{electric field} distribution at multiple AoAs for a $13\times 13$ Rotman lens.  From Fig.~\ref{fig:fig}(a), we see that the maximum power is concentrated on the central beam port, i.e., port 7, while a small portion of the power is spilled over to the neighboring ports. However, as \textcolor{black}{the AoA moves} towards $\phi=50^{\circ}$, \textcolor{black}{one can} observe not only \textcolor{black}{stronger EM} energy spillover but also reflections towards the opposite ports. This loss of focusing ability of lens arrays is one of their fundamental limitations and underlines the importance of careful circuit design and advanced signal processing.} A very recent contribution in this context is \cite{Power_leakage}, which examines the \textit{power leakage problem} (equivalent to the spillover problem mentioned before) in mmWave massive MIMO systems with lens arrays. The authors proposed a beam alignment precoding scheme to alleviate this inherent problem by developing a phase shifter network structure.

\textit{Open challenges:} It is an indisputable fact that the performance characterization of lens topologies in the presence of hardware imperfections requires synergies between communications engineers and microwave engineers. Unfortunately, these two communities have very often worked in isolation from each other, and this has created a critical knowledge gap.  \textcolor{black}{ There are many opportunities for research, and a} non-exhaustive list of open problems is the following:
\begin{enumerate}[a)]
\item The impact of the switching matrix is largely overlooked in the literature. In an ideal world, this matrix is binary and each row of it contains only one nonzero entry corresponding to the selected beam index. Yet, practical switches are not fully absorptive, which implies that energy is reflected back to the lens beam ports, whilst the poor isolation between switches causes energy leakage in the neighboring switches \cite{Tataria1}.

\item  The investigation of non-ideal mmWave RF components (e.g. mixers, local oscillators, power amplifiers) which induce \textcolor{black}{in-band and out-of-band} distortions is a very important topic since their aggregate impact can seriously undermine the theoretically predicted performance.
\end{enumerate}

\subsubsection{Physical implementation}
The physical implementation of \textcolor{black}{a communication system based on lens arrays} is a new topic and we will now pinpoint the most important advances.
We first recall that the two most popular approaches to implementing \textcolor{black}{EM} energy focusing using lens arrays are layered scattering and guided wave techniques \cite{lau_hum}. A contemporary overview of the different types of lens arrays and their key characteristics can be found in \cite{Lens_2017}. Also, we refer the readers to \cite{Rotman}, which meticulously covers Rotman lens-based MIMO systems with beam selection and digital beamforming.

A CAP-MIMO demonstrator at $10$\,GHz was first presented in \cite{brady_CAPMIMO} and later extended to multibeam operation at $28$\,GHz
in \cite{Sayeed2016, Sayeed2017}. In \cite{Chae_lens1}, a MIMO system at $77$\,GHz that uses different types of RF lenses was manufactured and measured. The authors also proposed a multivariance codebook quantization scheme to reduce the feedback overhead. The same group also developed a number of prototypes at $28$\,GHz using a hyperbolic, dielectric lens made from polyethylene for static and mobile applications \cite{Chae_lens2}. The authors in \cite{Laurinaho} presented a 2D beam steerable lens antenna prototype at $71$--$76$ GHz with a $64$-element feed antenna, that can deliver $700$\,Mbit/s throughput at an operating range of $55$\,m. Most recently, \cite{abbasi} synthesized and measured a $28$\,GHz lens array using constant dielectric material with antenna feeds for multibeam operation. This geometry was shown to systematically outperform a ULA and the Rotman lens solution of \cite{Fusco2} thanks to sharper \textcolor{black}{EM} focusing.

\section{Intelligent Reflecting Surface}
\label{sec:IRS}

While the great spectral efficiency gains offered by MIMO communication are well established \cite{BjornsonHS17}, there are doubts regarding the technology's ultimate cost and energy efficiency. In fact, it was shown in \cite{JR:EE_massive_MIMO_Kwan} that an exceedingly larger number of antennas is not the way to improve the energy efficiency (measured in bit/Joule) of future networks, as the total energy consumption increases linearly with respect to the numbers of RF chains required by the active components while the data rates only grows logarithmically.
There have been serval attempts to maximize the energy efficiency of MIMO systems by both optimizing the power allocation \cite{JR:Jie_Xu_MIMO_EE,JR:Nguyen_MIMO_FD_EE,Venturino2015a} and the network topology \cite{Bjornson2015a,Mohammed2014a,Bjornson2016aabb}. While these works find an optimal trade-off between data rates and energy consumption, the optimal design is often one with many hardware components and therefore high cost. Hence, the advantages of MIMO do not come for free and the performance improvements of the wireless technology might eventually saturate due \textcolor{black}{to} financial reasons.
Hence, the design of \textcolor{black}{spectrally- and energy-efficient communication system} with low hardware cost is of utmost importance for realizing economically sustainable  wireless communication networks \cite{Tombaz2011a,JR:QQ_5G_Green}. In this section, we explore one potential way to achieve that by using passive MIMO antennas in an architecture\footnote{\textcolor{black}{IRS is a passive surface that only reflects impinging RF signals generated from ambient transmitters. In contrast, Large Intelligent Surfaces and Holographic Massive MIMO are two names used for large active surfaces exploiting active MIMO antennas driven by active energy-hungry components \cite{Hu2018a}.}} known as \textcolor{black}{an} \emph{intelligent reflecting  surface (IRS)} \cite{Wu2018a,CN:Dongfang_IRS_GC,JR:Alex_IRS} or \emph{software-controlled metasurfaces} \cite{Liaskos2018a}.

\subsection{Motivation}

The roll-out of MIMO has fueled the development of high-speed wireless communication systems \cite{book:Vincent_book,JR:MIMO_Kwan}. However, the performance of a wireless system is still determined by its \textcolor{black}{channels}. Specifically, the \textcolor{black}{EM} waves radiated by a transmitter experience reflections, refraction, diffractions, and pathloss in the channel before reaching a receiver. Conventionally, the communication channel is treated as an uncontrollable environment which can be modeled probabilistically \cite{book:david_wirelss_com,s2010wireless}. In fact, most of the communication techniques developed in the literature (e.g. beamforming, diversity, channel coding) were designed to either counteract or exploit the effects of the channel without changing its behavior.  In contrast, the recently proposed IRS concept builds on manipulating the propagation of EM waves in a communication channel so as to improve the performance of communication systems. Specifically, an IRS is a metasurface consisting of a large set of tiny elements that diffusely reflects incoming signals in a controllable way. IRS builds on the classical concept of reconfigurable reflectarrays \cite{Shaker2014a} with the added requirement of having real-time reconfigurability and control.

\begin{figure}
\centering
\includegraphics[width=3.5  in]{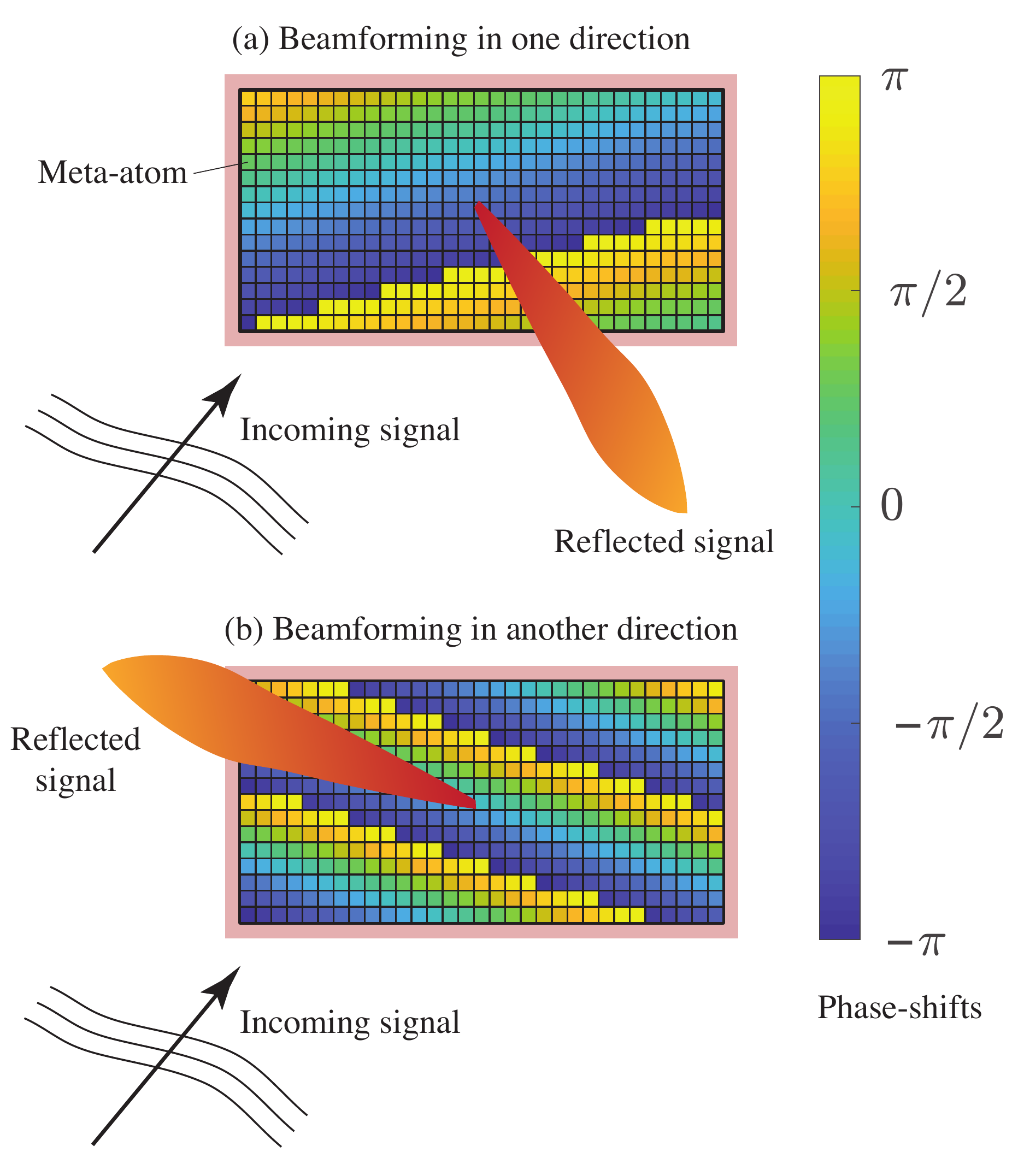}%
\caption{An IRS consists of many discrete meta-atoms of a sub-wavelength size, illustrated as colored squares. Each atom assigns a phase-shift to the incoming signal before it is scattered. The color of each atom represents its optimized phase shift values (indicated in the vertical color bar on the right hand side).   As illustrated in (a) and (b), different selections of the phase-shifts lead to beamforming from the IRS in different directions.} \label{fig_IRS_basics}
\end{figure}

Normally, a flat finite-sized surface reflects the incoming wave in the main direction determined by Snell's law \cite{Balanis2012a} but with \textcolor{black}{a beamwidth that is inversely proportional to} the size of the surface relative to the wavelength \cite{Larsson2019a}. While perfect specular mirror reflections often occur in the visible light range, that is typically not the case for \textcolor{black}{signals in cellular networks} which have \textcolor{black}{on the order of $10^4$ to $10^5$ times} longer wavelengths \cite{Larsson2019a}. The use of \textcolor{black}{metasurfaces} cannot change the reflection losses, but it can create anomalous \textcolor{black}{reflections} \cite{Liang2015a}, meaning that the main direction of the reflected signal can be controlled. This can be achieved by letting every point on the surface induce a certain \textcolor{black}{phase shift} to the incoming signal. Ideally, this should be done in a continuous way over the surface \cite{PhysRevX.8.011036}, but metasurfaces approximate this using many discrete ``meta-atoms'' of a sub-wavelength size that each induces a distinct phase-shift \cite{7792160}. Hence, an IRS is an array of meta-atoms that each scatter the incoming signals with a controllable phase-shift \cite{Wu2018a,JR:IRS_EE,CN:QQ_IRS_ICASP}, so that the joint effect of all phase-shifts is a reflected beam in a selected direction. This resembles beamforming from a classical phased array but with the main difference that the signal is not generated in the array but elsewhere. Fig.~\ref{fig_IRS_basics} illustrates how different phase-shift patterns among the meta-atoms lead to the incoming signal being reflected as a beam in different directions. Even if it tempting to view an IRS as a mirror, it actually behaves as a reconfigurable lens that can focus signals at points in the near-field or beamform signals towards points in the far-field \cite{Bjornson2020new}.

Unlike cell-free massive MIMO systems and cooperative relays, which also attempt to improve the propagation conditions by deploying active hardware components, an IRS is believed to only require a small operational power making it suitable for implementation in energy-limited systems. Besides, an IRS can operate naturally \textcolor{black}{in a full-duplex} manner
without the need of costly self-interference cancelation. For example, when using meta-atoms with the size $8\times 8$ mm, the energy consumption is only $125$ mW/$\text{m}^2$, which is considerably lower than for many existing wireless communication devices \cite{JR:Ian_meta_magazine}. Furthermore, an IRS can be of thin and conformable material, allowing for nearly invisible deployment on building facades and interior walls. Hence, once a conventional network has been deployed, one or multiple \textcolor{black}{IRSs} can be flexibly deployed to mitigate coverage holes that have been detected or to provide additional capacity in areas where that is needed.
\textcolor{black}{In fact, the IRS is not supposed to replace or compete with conventional massive MIMO technology, but rather complement it.} It is similar to the dish that is used in satellite receivers; it is a passive device that reflects signals to improve the SNR.

\begin{figure}
\centering
\includegraphics[width=3.5  in]{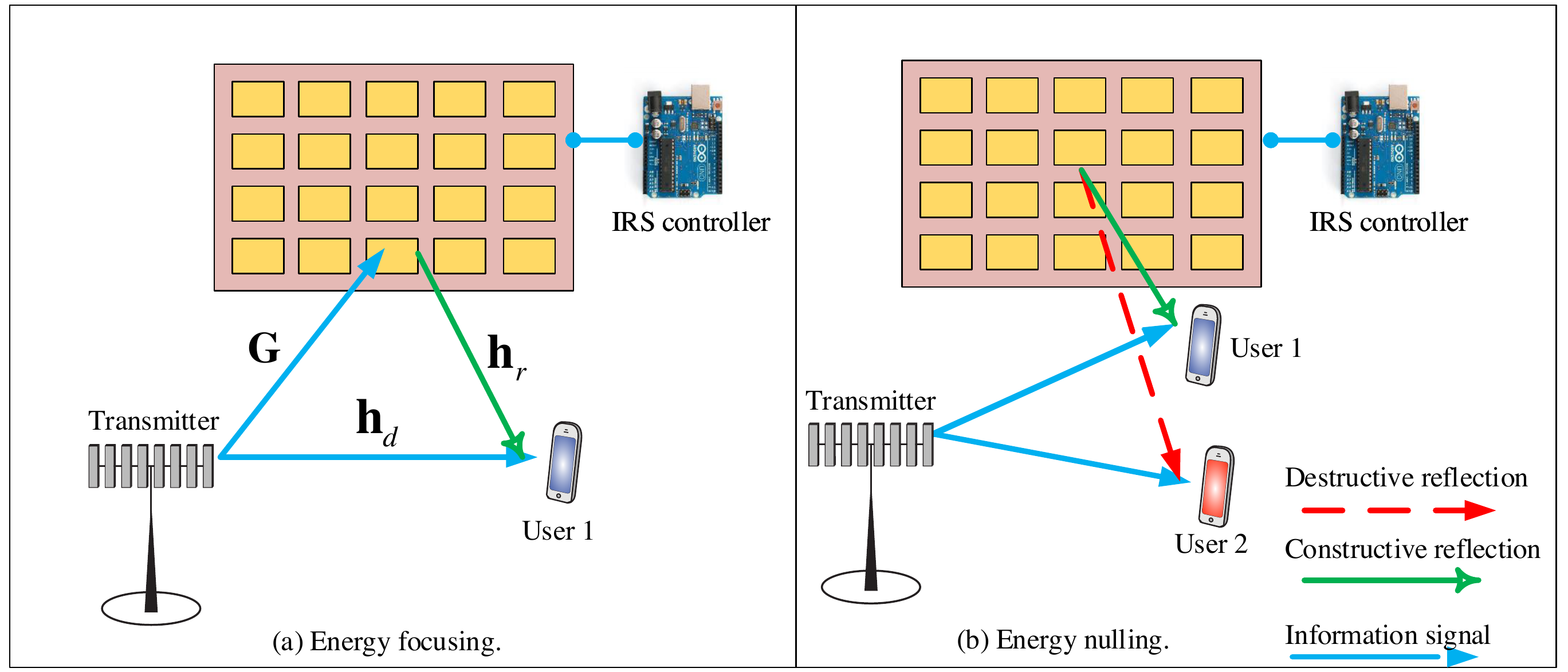}%
\caption{An IRS-enhanced  multiple antennas wireless communication system. } \label{fig_system_model_IRS}
\end{figure}

 In practice, the deployment of an IRS in conventional MIMO systems facilitates two types of beamforming which are illustrated in Fig.~\ref{fig_system_model_IRS}. In Fig.~\ref{fig_system_model_IRS}(a), there is one IRS deployed in a system assisting the communication between a multiple-antenna transmitter and a user. The information signal is radiated from the transmitter. A direct path may exist between the transmitter and the user for communication, and beamforming is performed at the transmitter to improve the signal reception at the user. Meanwhile, the information signal is also received by the IRS due to the broadcast nature of wireless channels and the IRS will reflect the signal. With the help of an IRS controller, the main direction of the reflection can be controlled. In particular, proper phase shifts are introduced on all the meta-atoms to deliberately create a coherent combination of their individually scattered signals, thereby creating a signal beam focused at the user. The larger the \textcolor{black}{surface} is, the narrower the beam will be. This strategy is known as \emph{energy focusing} \cite{Wu2018a,CN:QQ_IRS_ICASP,JR:QQ_IRS_Magaizne}.

On the other hand, if a direct path does not exist due to heavy shadowing or blockage, the transmitter should perform beamforming with respect to the IRS. Then, the IRS can act as a non-amplifying full-duplex relay which reflects and focuses the incident information signal to the UEs for assisting the end-to-end communication. In Fig.~\ref{fig_system_model_IRS}(b), we consider a scenario where a multiple-antenna transmitter serves user $1$ in the \textcolor{black}{presence} of user $2$. We assume the two UEs have different security clearance levels where the message of user $1$ should not be decodable at user $2$.  In this situation, destructive reflection can be performed at the IRS, by adjusting the phases of the scattered signals to null out the signal at user $2$. This strategy is known as \emph{energy nulling} \cite{Wu2018a,CN:QQ_IRS_ICASP,JR:QQ_IRS_Magaizne}. \textcolor{black}{By exploiting these two principles, it is expected that IRSs have wide applications \textcolor{black}{in various} communication systems involving interference management, coverage extension, \textcolor{black}{and capacity improvement, such as in} wireless-powered communication systems, cognitive radio networks, physical
layer security systems, etc.}

\subsection{Signal Model of IRS}
The amalgamation of IRS and conventional communication systems has introduced a new paradigm for the design of energy-efficient communication. In this paper, we focus on the point-to-point communication system in Fig.~\ref{fig_system_model_IRS}(a) for the illustration of the signal model. There is a transmitter equipped with $M$ antennas serving a single-antenna user. In particular,  an IRS \textcolor{black}{consisting} of $N$ meta-atoms elements is deployed to assist the end-to-end communication. Besides, an IRS controller is adopted to control each meta-atom such that the phase of the scattered incident signal can be dynamically adjusted to achieve different purposes. Assuming deterministic flat-fading channels, the signal received at the single-antenna user is
\begin{equation} \label{eq:system-model-IRS}
y= (\underbrace{\overbrace{ \mathbf{h}^{\Ttran}_r\boldsymbol{\Theta} \mathbf{G}}^{\text{Reflected path}} + \overbrace{ \mathbf{h}^{\Ttran}_d}^{\text{Direct path}}}_{\text{Composite channel}})\mathbf{w}x + z,
\end{equation}
where $x\in \mathbb{C}$ and $\mathbf{w}\in \mathbb{C}^{M\times 1}$ are the unit-power information symbol and beamforming vector from the transmitter for the user, respectively. Furthermore, $\mathbf{h}_d\in \mathbb{C}^{M\times 1}$, $\mathbf{h}_r\in \mathbb{C}^{N\times 1}$, and  $\mathbf{G}\in \mathbb{C}^{N\times M}$ denote the channels of the transmitter-to-user, IRS-to-user\textcolor{black}{,} \textcolor{black}{ and} transmitter-to-IRS links, respectively. \textcolor{black}{The $N$-by-$N$ diagonal matrix $\boldsymbol{\Theta}\in \mathbb{C}^{N\times N}$ contains complex exponentials, $e^{j\theta_n},\forall n\in\{1\ldots,N\}$, on the diagonal, where $\theta_n\in[0,2\pi]$ is the phase-shift introduced at the $n$th meta-atom of the IRS.}
Finally, $z \sim \CN(0,\sigma^2)$ denotes the independent noise at the receiver.

\textcolor{black}{To obtain} the system model\footnote{This system model can be used for many different purposes. We only exemplify the use of it for information transfer, but we stress that wireless power transfer, cognitive radio, information nulling, etc. can also be considered in  future works.} in \eqref{eq:system-model-IRS}, it is assumed that the delay spread of the reflected path is approximately the same as the delay spread of the direct path, which is valid if the IRS is placed close to either the transmitter or the user. The channels $\mathbf{h}_d$, $\mathbf{h}_r$, $\mathbf{G}$ can be modeled as conventional MIMO channels, and each might consist of multiple paths. The antenna gains at the transmitter, receiver, and each meta-atom are also included in the channels to keep the notation simple.

\subsubsection{Rate Optimization}

If the receiver knows the composite channel, its achievable rate is $\log_2(1+\mathrm{SNR})$, where the  signal-to-noise ratio (SNR)  at the user is
\begin{align}\label{SectionII:receivedpower}
\mathrm{SNR} = \frac{|{( \mathbf{h}^{\Htran}_r\boldsymbol{\Theta} \mathbf{G}+\mathbf{h}^{\Htran}_d)\mathbf{w} }|^2}{\sigma^2}.
\end{align}
To maximize the rate, the transmit beamforming vector $\mathbf{w}$ and the phase-shift matrix $\boldsymbol{\Theta}$ can be jointly optimized to maximize the SNR. This can be mathematically formulated as the following optimization problem:
\begin{eqnarray}
 ~~\underset{{\mathbf{w}, \boldsymbol{\Theta}}}{\mathrm{maximize}}  ~~~&&| (\mathbf{h}^{\Htran}_r{\boldsymbol\Theta} \mathbf{G}+\mathbf{h}^{\Htran}_d )\mathbf{w}|^2  \label{eq:obj}\\
\mathrm{\,\,subject\,\, to}~~~&& \|\mathbf{w}\|^2\leq P_{\max},\label{power:constraints}  \\
&& 0\leq \theta_n \leq 2\pi, \quad \forall n=1,\ldots, N, \label{phase:constraints}
\end{eqnarray}
where $P_{\max}$ in \eqref{power:constraints} denotes the maximum transmit power budget of the transmitter and \eqref{phase:constraints} is the constraint on the phase introduced by each reflection element.
Although the constraints span a convex feasible set, the objective function in \eqref{eq:obj} is non-concave due to the coupling between $\mathbf{w}$ and $\boldsymbol{\Theta}$. \textcolor{black}{Another big challenge is the constant modulus elements of $\boldsymbol{\Theta}$.}
For $M\geq2$, there is generally  no systematic approach for solving such non-convex optimization problems optimally and efficiently. In some cases,  brute force approaches are needed to obtain the globally optimal solution which incurs a prohibitively high computational complexity even for moderate-sized systems.  To strike a balance between system complexity and performance, different suboptimal approaches (e.g. alternating optimization, semidefinite relaxation, successive convex approximation, manifold optimization, etc.) have been proposed in the literature to obtain a computationally efficient solution \cite{Wu2018a,CN:Dongfang_IRS_GC,CN:QQ_IRS_ICASP,JR:QQ_IRS_Magaizne,JR:Alex_IRS}.

\begin{figure}[t]
\centering
\begin{subfigure}{.45\textwidth}
\centering
  \includegraphics[width=3.5 in]{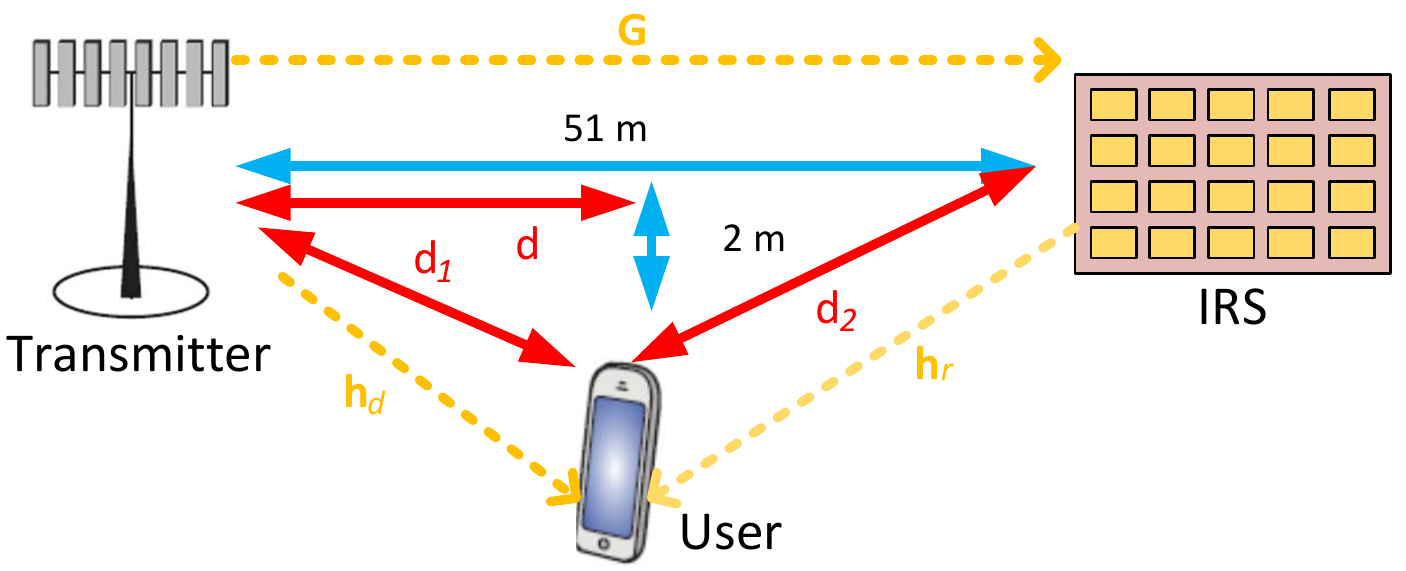}%
\caption{\footnotesize Simulation setup.}\label{fig_simulation_IRS_setting}
 \end{subfigure}\\
 \begin{subfigure}{.5\textwidth}
\includegraphics[width=3.5 in]{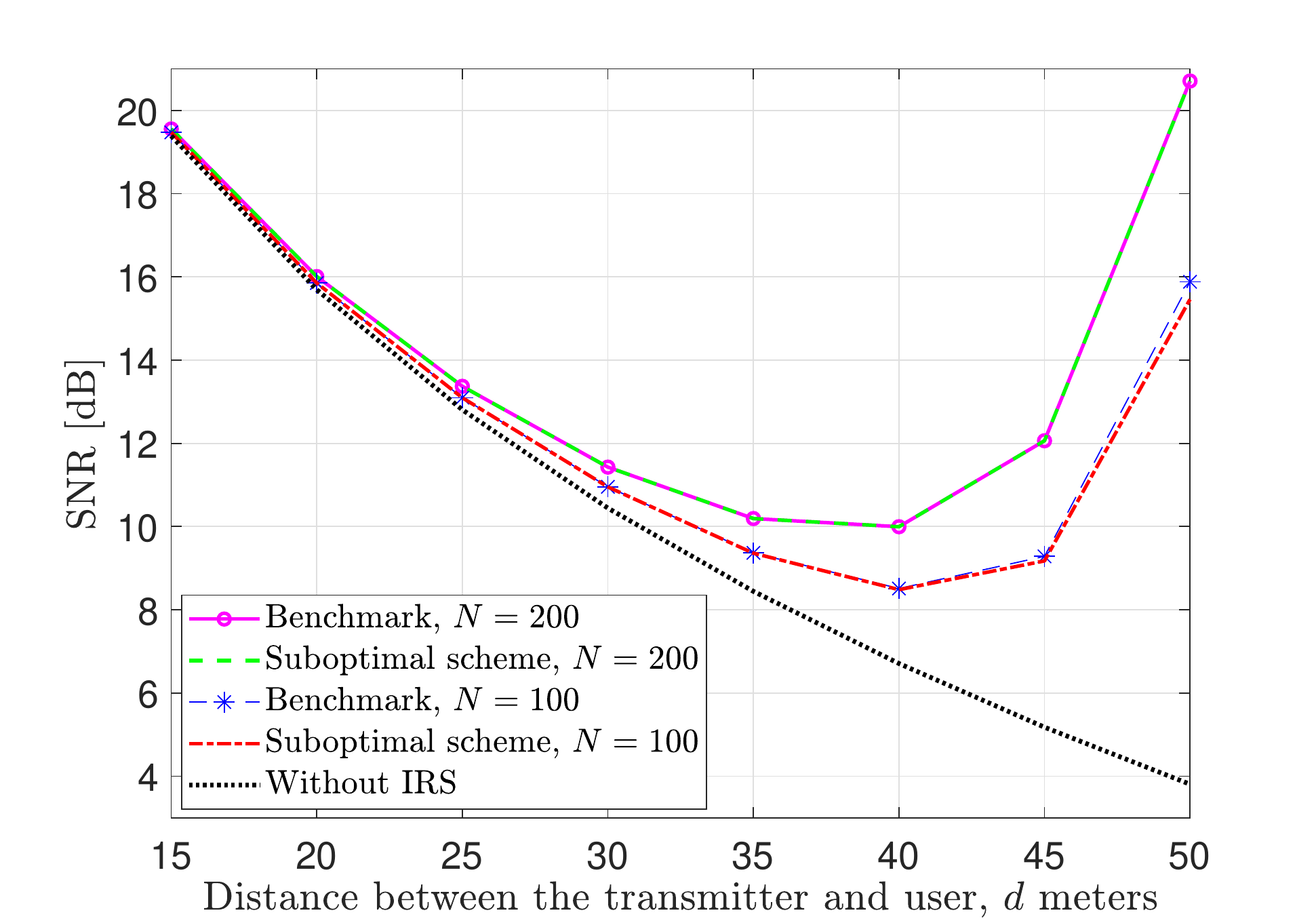}
\caption{\footnotesize SNR versus the horizontal distance $d$ between the transmitter and the user for various schemes}
\label{fig_simulation_IRS_curve}
 \end{subfigure}
 \caption{\footnotesize \color{black}SNR versus the horizontal distance $d$ between the transmitter and the user for various schemes. It is assumed that  the transmitter-to-IRS channel is
a free-space LoS link while the IRS-to-user
channels are modeled by independent Rayleigh fading with a pathloss
exponent of $3$.}\label{fig_simulation_IRS}
\end{figure}

In Fig.~\ref{fig_simulation_IRS}, we provide simulation results illustrating the performance gain brought by the deployment of an IRS in a MIMO system. We follow a similar system setting as in \cite{Wu2018a}. We assume
that there is an IRS located at $51$\,m away from a MIMO transmitter equipped with $M=2$ transmit antennas. There exists a user located on a horizontal line which is parallel to the one connecting the transmitter and the IRS.As illustrated \textcolor{black}{in Fig.~\ref{fig_simulation_IRS} (a), the horizontal distance from the transmitter to the user is a variable $d$ and the vertical distance between the user and the horizontal line connecting the BS and IRS is $2$\,m.} We consider a $1$ MHz bandwidth, for which the transmit power is $0$ dBm and the noise power is $-110$ dBm. There are $N$ meta-atoms forming a rectangular array\footnote{\color{black}In practice, the position arrangement of atoms at the IRS may have some impact on the system performance when the communication distance between the transmitter/user is short compared to the physical size of the IRS. However, such near-field effects have virtually no impact for the values of $N$ and distances considered in this simulation \cite{Bjornson2019e}.} at IRS with  $a$ rows and  $b$ columns such that  $a\times b=N$.
Fig.~\textcolor{black}{\ref{fig_simulation_IRS} (b) shows the SNR that is achieved with different transmission schemes as a function of the distance $d$.} We compare the SNR that is achieved by three schemes: 1) benchmark scheme, 2) suboptimal scheme, 3) baseline without an IRS. The results of the benchmark and suboptimal schemes are obtained by solving the optimization problem in \eqref{eq:obj} via semi-definite relaxation (SDR), leading to an upper bound, and the SDR with Gaussian randomization, respectively. First, it can been seen from Fig.~\textcolor{black}{\ref{fig_simulation_IRS} (b)} that the suboptimal scheme achieves a close performance compared to the upper bound. In other words, the suboptimal scheme represents the actual achievable SNR by deploying an IRS. Second, it is obvious that the introduction of an IRS can substantially increase the SNR compared to the baseline without an IRS. The SNR gains are largest when the user is close to the IRS, which is logical since otherwise the reflected path would be too weak make a true contribution in the numerator of \eqref{SectionII:receivedpower}.

\subsection{Open Research Problems}
The introduction of an IRS into a traditional communication system revolutionizes the
design of beamformers and network topologies. In the following, we discuss some research challenges for IRS-assisted MIMO communication systems.

\subsubsection{Channel Estimation}

The performance of an IRS depends on its beamfocusing capability which relies on the availability of CSI at the IRS controller. In other words,  we need to acquire the information about the channels  $\mathbf{h}_r$, $\mathbf{h}_d$, and $\mathbf{G}$ in order to properly select $\boldsymbol{\Theta}$ and $\mathbf{w}$.  In general, the transmitter-to-user link, $\mathbf{h}_d$, can be obtained by applying traditional channel estimation strategies based on pilot transmission. In contrast, the channel estimation for the transmitter-to-IRS and the IRS-to-user links is more challenging due to the following three reasons. First, an IRS consists of passive elements which cannot initiate transmission to facilitate accurate channel estimation. Second, although the IRS controller may be equipped with a simple communication module for exchanging wireless control signals between the IRS and the transmitter, its limited computational ability may introduce an exceedingly long delay for estimating both $\mathbf{G}$ and $\mathbf{h}_r$. Third, in order to achieve a reasonable system performance, a large aperture with a large number of meta-atom elements is required \cite{Bjornson2019IRS}. For example, there can be $15,625$ meta-atoms in a $1\times 1$\,m IRS, if each one is $8\times 8$ mm as was considered in \cite{JR:Ian_meta_magazine} for a carrier frequency of $5$\,GHz. Performing channel estimation for such a high-dimensional channel would put heavy signal processing burdens and energy consumption at wireless transceivers. Hence, there is an emerging need for the design of low-cost channel estimation algorithms for IRS-supported communication, which might be achieved using parameterizable channel models or sparsity \cite{He2019a}. Also, \textcolor{black}{the} beamspace approach, described in Section~\ref{section:beamspace}, might be a key part of the solution. Besides,  there are some initial attempts in the literature for addressing the channel estimation problem in IRS-assisted systems. For example, in \cite{JR:Compressive_sensing_CSI}, compressive sensing-based channel estimation was studied for IRS-assisted mmWave systems. In \cite{JR:IRS_CSI}, a practical transmission protocol was proposed to execute channel
estimation and reflection optimization successively. In \cite{JR:IRS_CSI_three_phases}, a tailor-made three-phase pilot-based channel estimation framework was designed to estimate the overall uplink channels in an IRS system. However, the related energy costs for channel estimation and signaling overhead in these works are not taken into \textcolor{black}{account, which somewhat conflicts with} the purposes of applying IRSs. Overall, designing a practical channel estimation for IRS systems with \textcolor{black}{low energy consumption, low signaling overhead, and low computational complexity is still an open problem which requires dedicated research efforts for further investigation.}

\subsubsection{Controlling IRSs}
\textcolor{black}{Despite various preliminary research works have been conducted to
unlock the potential of IRSs, controlling IRSs in practical systems is challenging. In fact, in order
to fully exploit the performance gain of IRS systems, a smart controller should be installed at the
IRS which controls the reflection amplitude and phase of the reflected signals. Besides, the smart
controller should communicate with the transmitter to facilitate the estimation of CSI and for real-time adaptive beamforming. Furthermore, certain time synchronization
control between a transmitter and an IRS is needed. However, practical control protocols for
smart controllers are still unavailable in the literature. \textcolor{black}{It is} still unclear if the IRS
should be controlled via a separate wireless link or via dedicated time slots. One may \textcolor{black}{follow a
similar approach/ protocol as narrowband Internet of things} systems for handling controlling signals\footnote{Note however that although the hardware of meta-atom-made IRS has been realized by some existing proof-of-concept prototype of
IRS systems, e.g. \cite{JR:IRS_prototype}.}. Yet, if
the smart controller is powerful enough for handling complicated real-time controlling protocols
between the transmitter and the IRS, the associated energy consumptions of the IRS will become
a concern of system performance as IRSs are supposed to be energy-limited devices.}

\subsubsection{Hardware Impairments}
The system models for IRS-assisted communication have been developed based on
the assumption of perfect hardware. However, both IRSs and IRS controllers are preferably fabricated using low-cost components which will be subject to hardware impairments. For example, the phase-shifting capability might have limited resolution, such as only two states \cite{Liang2015a}. A preliminary study on the impact of finite resolution phase shifting is provided in \cite{CN:QQ_IRS_ICASP}. Thorough performance analysis of IRS-assisted systems taking into account the impact of mutual coupling, phase noise, and other hardware impairments is desired in future work.

\subsubsection{Deployment of IRS}

The position of an IRS is crucial for achieving a useful performance improvement in IRS-assisted communication systems, as shown in Fig.~\ref{fig_simulation_IRS}. The SNR of the reflected path is proportional to the product of the pathlosses between the transmitter and the IRS and between the IRS and the receiver \cite{Larsson2019a}, which is why physically large surfaces are needed for an IRS to beat competing range-extension technologies such as relays \cite{Bjornson2019IRS}. Under optimized transmission, the SNR grows with the square of the surface area, since the area first determines \textcolor{black}{how large a fraction of the transmitted power that reaches the IRS} and then determines how narrowly it can be beamformed towards the receiver \cite{Bjornson2019e}.
Hence, if a large IRS is placed at a location with clear \textcolor{black}{LoS} with respect to the transmitter and close distance to the receiver, it can efficiently increase the SNR. However, such kind of deployment may not work well for helping the transmitter to convey multiple data streams for multiple UEs. In fact, the single strong LoS  path  between the IRS  and the transmitter might result in a low-rank MIMO channel which only offers a limited spatial multiplexing gain. To achieve a multiplexing gain, one should consider deploying multiple IRSs in the system to artificially create sufficient numbers of controllable ``scatterers". Yet, the optimal positioning of IRSs for maximizing the total system capacity  is an open problem. {Besides, the problem of controlling multiple IRSs and the associated channel estimation problem for joint reflection remains unsolved in the literature.}

\subsubsection{Identifying a ``Killer Application''}

There is a series of survey papers that speculate on how an IRS can be used in future networks \cite{JR:Ian_meta_magazine,JR:QQ_IRS_Magaizne,Renzo2019a,Bjornson2019d}, but there is a need to demonstrate a ``killer application'' or a metric for which IRS-aided transmission \textcolor{black}{makes} a paradigm shift in performance. Massive MIMO provides orders-of-magnitude improvement in spectral efficiency in sub-6 GHz bands and enables unprecedented data rates in mmWave bands. What will be the corresponding key benefit of IRS? Energy focusing and energy nulling, as illustrated in Fig.~\ref{fig_system_model_IRS}, can be also achieved with conventional beamforming methods, but \textcolor{black}{likely would result} in higher implementation cost. Similarly, range extension is the classical use case of relays, which already have a low cost  \cite{Bjornson2019IRS}. A relay can be much smaller than an IRS since its transmit power determines the signal amplification it provides, not the physical size as is the case of the IRS. The ability to control the propagation environment is conceptually appealing but must be associated with a practical performance gain. \textcolor{black}{The research community needs} to think outside the box to find the right context in which IRS-supported systems will prevail.

\section{MIMO Meets Others}
\label{sec:MIMO-meets-others}

The evolution of cellular technology has focused on creating a single system that can simultaneously support all communication applications. However, some important future applications (e.g. augmented reality (AR), virtual reality (VR), holography, ultra-reliable coverage) have so stringent requirements that the \textcolor{black}{network needs} to reconfigure itself to support these applications. It is envisioned that MIMO technology will continue to be a main driving force for the development in beyond-5G networks. In the following, we provide a brief discussion on how other communication technologies would complement MIMO for fulfilling upcoming stringent quality-of-service (QoS) requirements set by potential future use cases.

\subsection{Unmanned Aerial Vehicle-Based MIMO Communication}\label{se:UAV-MIMO}

\begin{figure*}
\centering
\includegraphics[width=5  in]{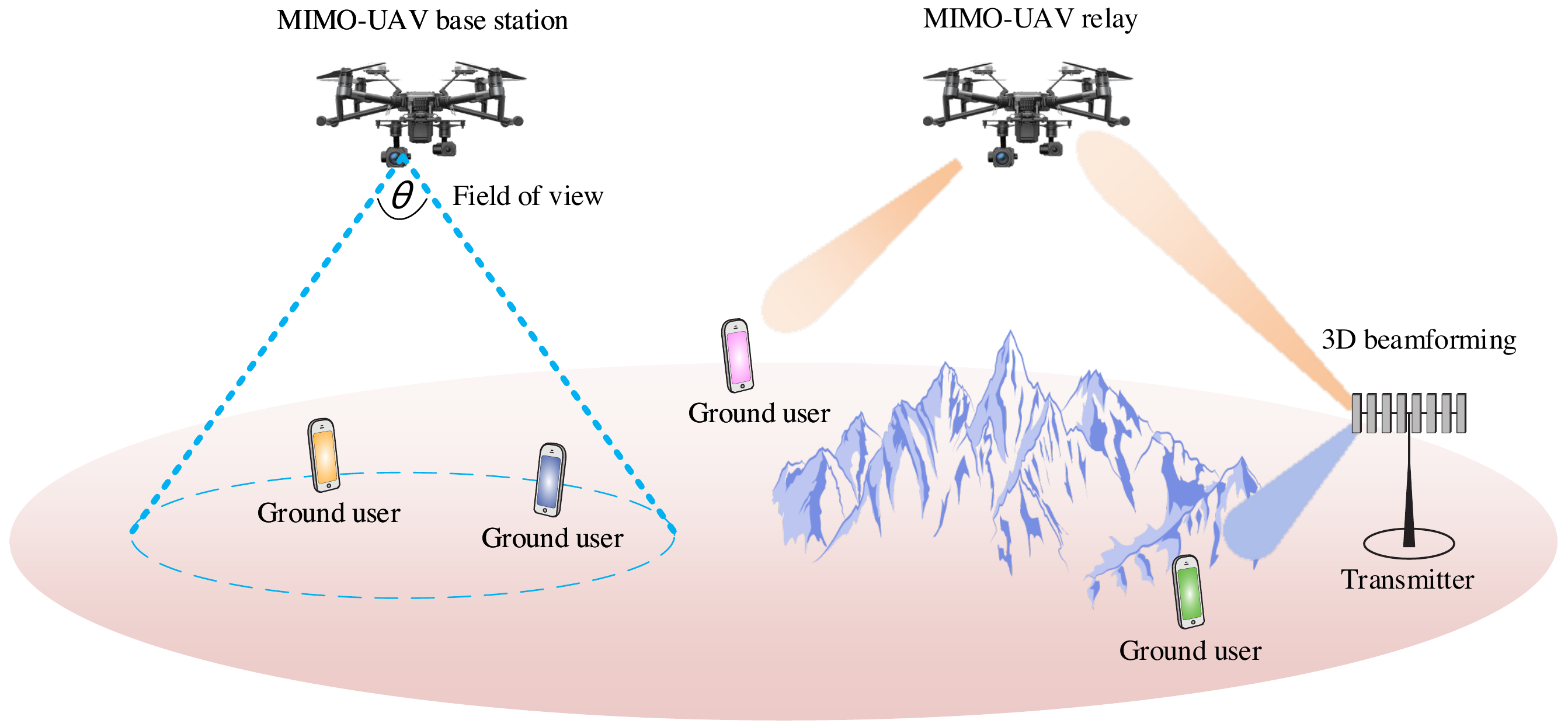}%
\caption{Different aspects of a MIMO-UAV assisted wireless communication system.} \label{fig_system_model_UAV}
\end{figure*}

Although \textcolor{black}{massive MIMO} can improve the SNR of a communication link proportionally to the number of antennas, there are harsh physical environments where this is not enough to provide decent coverage and capacity.  For example, if heavy shadowing exists between a transmitter and a receiver due to blockage, the user device might not be able to connect to the network at all. Although \textcolor{black}{cell-free massive MIMO}, described in Section~\ref{section:cell-free}, can be used to shorten distances between transmitter and receivers, this might not be sufficient to combat the shadowing due to the physical and financial constraints on where the fixed network infrastructure can be deployed.  In fact,  deploying a terrestrial infrastructure might be neither cost-effective nor feasible in practical cases such as in complex terrains, private areas, or remote areas. Also, on some occasions, terrestrial wireless networks \textcolor{black}{may malfunction} due to natural disasters, power outages, maintenance, etc. To handle these issues, aerial communication systems based on UAVs is regarded as a promising new paradigm to facilitate fast and highly flexible deployment of communication infrastructure due to their high maneuverability \cite{JR:QQ_magazine_UAV,JR:Yong_magazine_UAV,JR:QQ_UAV_OFDMA,JR:Yan_UAV_TCOM,CN:Ruide_GC}. In practice, UAVs equipped with MIMO communication modules can act as aerial base stations or aerial relays to assist communication. Fig.~\ref{fig_system_model_UAV} shows two common usages of UAVs. On the right-hand side of the figure, there is a blockage between a MIMO transmitter and a ground user. Then,  a UAV can be deployed as an aerial relay for establishing a reliable communication link between the desired transceivers. It can either use traditional relaying protocols or, potentially, the new IRS approach described in Section~\ref{sec:IRS}.
On the left-hand side of Fig.~\ref{fig_system_model_UAV}, when communication infrastructure is out of service or unavailable, a UAV can be dispatched to serve as an aerial base station to create a temporary communication hotspot for multiple ground UEs. Unlike conventional ground base stations and relays deployed at fixed locations, the high mobility of UAVs introduces additional spatial degrees of freedom in resource allocation for improving system performance. In particular, when the locations of the UEs are known, UAVs can adapt their trajectory for flying  close to a region with dense UEs to offer efficient communication services.

Despite the promising benefits brought by UAVs \cite{JR:QQ_magazine_UAV,JR:Yong_magazine_UAV}, the integration of conventional communication systems and \textcolor{black}{UAVs have} imposed new challenges to researchers for the design of efficient MIMO communication systems. First, the unique constraints on size, weight, and energy consumption of UAVs are the major obstacles in applying conventional communication theories for improving the system performance.  In practice, small UAVs (less than $25$ kg) are commonly deployed due to safety reasons \cite{JR:Yong_magazine_UAV}. Under that weight restriction, UAVs can only carry a light load to achieve high mobility and low energy consumption. In particular, the limited surface area, battery capacity, and computational capability do not facilitate the implementation of \textcolor{black}{massive MIMO} on UAVs. Second, in MIMO-UAV communication systems, the three types of channels (i.e., air-to-ground, ground-to-air, and air-to-air) are LoS dominated. Although the LoS channel characteristics facilitate the establishment of strong desired communication links, they are also vulnerable to multi-cell interference or potential eavesdropping \cite{CN:Cui_UAV_TVT,CN:Dongfang_UAV_GC,CN:Xiaofang_UAV,JR:Xiaofang_UAV_security_magazine}. While pilot contamination might be less of a problem for ground-to-ground than initially believed \cite{BjornsonHS17}, it is a major concern in MIMO-UAV communications.
Furthermore, the low-rank nature of \textcolor{black}{LoS} MIMO channels also offers limited spatial multiplexing capabilities for carrying multiple data streams. Thus, novel MIMO communication techniques enabling efficient interference management and exploiting multiplexing gains are needed, potentially by exploiting multiple UAVs to create distributed arrays with a large aperture. Third, the performance of a MIMO-UAV depends on its trajectory and resource allocation design. In general, since the channels in MIMO-UAV systems are LoS dominated, there is a non-trivial coupling between the trajectory of UAVs, location of UEs, and  beamforming vectors design via trigonometric geometry. The joint design of trajectory and resource allocation generally leads to non-convex problems and obtaining a globally optimal solution is challenging if not impossible. Up to now, there only exists one algorithm for achieving a globally optimal solution of the joint design problem for the case of a single-antenna UAV \cite{JR:Yan_UAV_TCOM}. As a result, the use of advanced global optimization techniques is needed to solve the design optimization problems of MIMO-UAV to unleash its full potential for performance improvement.

There are many open problems related to realizing efficient MIMO-UAV systems:

\subsubsection{3D Beamforming}

Traditionally, directional high-gain antennas with predefined antenna patterns are deployed to focus the signal energy onto the ground-based coverage area and simultaneously \textcolor{black}{reduce} inter-cell interference. Although this kind of deployment has worked well for ground UEs in past decades \textcolor{black}{(until massive MIMO became a standard feature of 5G)}, it is not a viable option for UAVs. In particular, a fixed beam pattern is only suitable for fixed deployments with a 2D distribution of UEs, while a UAV is highly mobile and \textcolor{black}{serves} a 3D distribution of UEs. Hence, 3D beamforming is suitable for UAVs \cite{JR:Yong_magazine_UAV_cellular,JR:3D_beamforming}. By deploying a two-dimensional array at the bottom of a UAV and controlling the magnitudes and phases of each antenna element, adaptive beamforming can be achieved towards different locations on the ground. Unfortunately, the beamwidth is limited by the array's aperture \cite[Sec.~7.4]{massivemimobook} and thus a limited-sized UAV cannot achieve particularly narrow beams. However, zero-forcing and other interference mitigation methods can still be implemented. Efficient algorithms for joint 3D beamforming at the base station and trajectory design of the UAVs need to be developed.

\subsubsection{\textcolor{black}{UAV Cooperation} (Distributed MIMO)}

Due to the physical constraints on weight, size, and energy consumption of UAVs, only a small number of onboard antennas can be \textcolor{black}{equipped on} every single UAV which limits the MIMO gains. Nevertheless, the angular field of view (FoV) of \textcolor{black}{a UAV} is usually small to ensure high SNR at the desired ground UEs, as shown in the left part of Fig.~\ref{fig_system_model_UAV}. Yet, the small FoV can only provide limited coverage for communications.  For example, when there is only one UAV serving a communication network having a large coverage area,  a UAV may need to \textcolor{black}{fly back} and forth in the system to satisfy the QoS requirements of all the UEs. This may introduce an exceedingly long delay and energy consumption in the system. Alternatively, the UAV can fly to a higher altitude \textcolor{black}{to create a large coverage area} but that is associated with lower SNRs. One effective solution is to deploy multiple UAVs for improving system performance.  In particular, a large service area can be divided into several small clusters,  each of which is served by a small number of UAVs. Meanwhile, if there is high-rate communication among UAVs via air-to-air communication links, distributed MIMO arrays can be formed by sharing antennas among all cooperative UAVs in each cluster for better interference mitigation and information transmission \cite{barriac_mudumbai_madhow,mudumbai_firstDist,brown_poor,goguri_dist}.  Depending on how this is implemented, methodology from the \textcolor{black}{cell-free massive MIMO} literature can be reused to achieve a lean and scalable architecture. In summary, a thorough study on user clustering and UAV cooperation algorithms \textcolor{black}{is needed to realize} the performance gain that can be achieved via joint transmission from multiple UAVs.

\begin{figure*}
\centering
\includegraphics[width=5in]{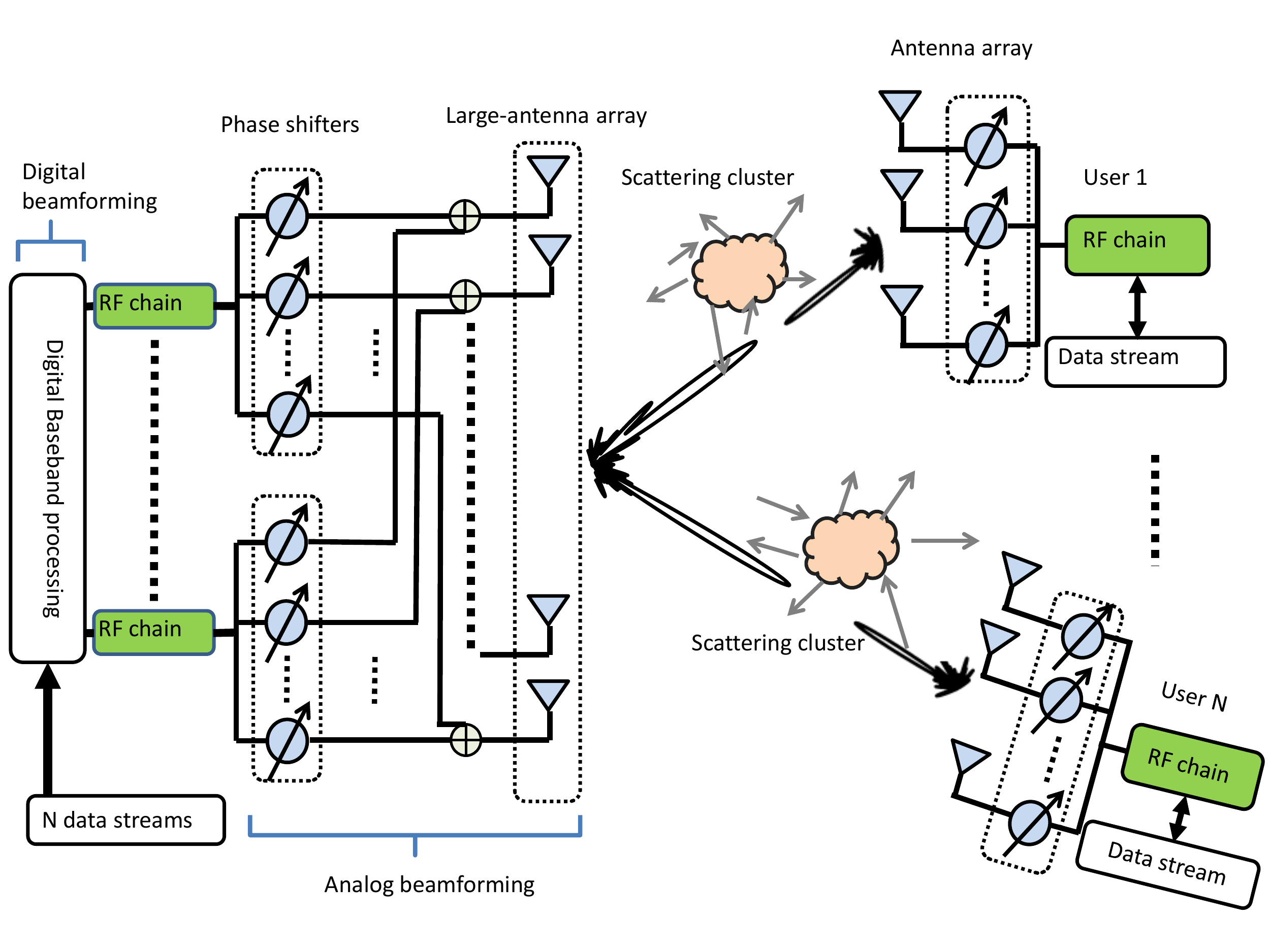}%
\caption{A mmWave communication system with hybrid analog-digital transceivers \cite{JR:Lou_mmwave}.} \label{fig_system_model_mmwave}
\end{figure*}

\subsection{MIMO for Sub-Terahertz Communications}

Each new network generation usually requires new spectrum bands to enable deployment on top of legacy networks. Since the spectrum is scarce in sub-6 GHz bands, 5G will rely on a combination of sub-6 GHz spectrum that can provide wide-area coverage and mmWave spectrum for high capacity in hotspots \cite{Parkvall2017a}. The main reason for the exploration of mmWave spectrum is the relatively wide bandwidths that are available, while the main drawbacks are the more complicated hardware design and propagation conditions \cite{rappaport2013millimeter}. The pathloss in free space is the same at any band but the antenna size shrinks with an increasing carrier frequency, thus MIMO arrays are necessary for mmWave bands to achieve the same antenna aperture as in legacy systems. The implementation of compact MIMO arrays is one of the challenges that \textcolor{black}{is tackled} in 5G literature \cite{JR:mmWave_MIMO_systems,Xiao2017a,JR:hybrid_MIMO_magazine}. Furthermore, mmWave signals  interact with objects in the propagation environment in a less favorable way, often limiting the coverage area to LoS scenarios \cite{Bjornson2019a}.

The first release of 5G supports the spectrum range \textcolor{black}{from $450$ MHz to $52.6$ GHz} \cite{Parkvall2017a}. The range might increase in future releases but it is fair to say that 5G is a technology for the sub-100 GHz \textcolor{black}{frequency range}. To identify new spectrum bands for beyond-5G \textcolor{black}{networks, systems} will have to move beyond the $100$ GHz barrier. There are extremely wide bandwidths available above that barrier\textcolor{black}{, at least 50\,GHz in the range of $90$--$200$\,GHz \cite{Saad2018a} and $100$\,GHz in the range of} $220-320$\,GHz \cite{Shams2016a}. Around $21$ GHz of this spectrum is currently open for unlicensed use in the USA \cite{Rappaport2019a}. \textcolor{black}{The World} Radiocommunication Conference 2019 has allocated $137$ GHz for the land-mobile and fixed services applications \cite{WRC19-Acts}. Although the concerned bands are, formally speaking, mmWave bands, it has become popular to call \textcolor{black}{bands in} the $100$--$300$ GHz range \emph{sub-THz bands} to distinguish them from the mmWave bands considered in 5G \cite{Rappaport2019a}.

The research into sub-THz communications is still in its infancy, but in comparison with 5G, \textcolor{black}{one can} expect even more directive transmission and limited range, with channels only consisting of LoS paths and \textcolor{black}{possibly} a few single-bounce reflections. On the other hand, with a bandwidth \textcolor{black}{of $100$ GHz}, the data rates can reach $1$ Tbit/s at the cell center and $100$ Gbit/s at the cell edge \cite{Rappaport2019a}. Reaching such extreme data rates is the main reason for considering \textcolor{black}{sub-THz frequencies}. It might become suitable for fixed outdoor installations with high-gain antennas (e.g. fixed wireless access or fronthaul/backhaul links \cite{Edstam2017a}) and short-range transmission to UEs that are mobile but move very slowly \cite{Faisal2019a}. There is also a potential of using sub-THz signals for sensing, imaging, and positioning \cite{Rappaport2019a}.

In the remainder of this section, we will briefly describe some of the MIMO-related challenges that will appear in sub-THz communications.


\subsubsection{Hybrid Beamforming}

The design of mmWave communications has become almost synonymous with hybrid analog-digital beamforming implementations, where a low-dimensional digital beamformer is combined with analog beamforming; \textcolor{black}{the latter may, for example, be} implemented using phase shifters as shown in Fig.~\ref{fig_system_model_mmwave}. \textcolor{black}{Hybrid beamforming} represents a compromise between hardware complexity and beamforming flexibility that simplifies the initial transceiver design. However, for the mmWave bands considered in 5G, hybrid beamforming \textcolor{black}{may be} a temporary solution that is applied until digital transceivers can be built \cite{Bjornson2019d}. In fact, there are already digital testbeds for the $28$ GHz band \cite{Tawa2018a}. When moving to the sub-THz band, it will be natural to once again begin with building analog or hybrid transceivers \cite{Akyildiz2016a,Faisal2019a}. Each antenna element will be so small that one cannot \textcolor{black}{yet} place a dedicated RF chain behind it \cite{Rappaport2019a}. By having a smaller number of RF chains connected to the antenna elements, the maximum beamforming gain can still be obtained \textcolor{black}{but} the beamforming design becomes more cumbersome.

Many of the same challenges that have been previously tackled in the mmWave literature \cite{JR:mmWave_MIMO_systems,Xiao2017a,JR:hybrid_MIMO_magazine} need to be revisited, under partially different conditions: a) More antenna elements per RF chain; b) Increasing frequency-selectivity in the channels due beam-squinting over wide bandwidths; c) Unknown propagation behaviors make it harder to use parameterized models for channel estimation; and d) Hardware impairments such as phase noise and non-linearities become increasingly influential on the performance \cite{moghadam_energyeff}.
It is plausible that low-resolution hardware must be utilized to achieve a cost and energy efficient implementation \cite{JR:Heath_mmwave_ADC,JR:David_Love_mmwave_ADC,zhang2017performance,zhang2016spectral}, and there is a hope that the resulting distortion will (partially) fall into the null-space of the desired communication links \cite{Bjornson2014a,Rappaport2019a,zhang2018mixed,Bjornson2019dist}.
It is necessary to first determine how the channel and hardware constraints will affect sub-THz bands and go back to the hybrid beamforming literature and determine which features remain. The beamspace approach, described in Section~\ref{section:beamspace}, is a suitable methodology when designing the signal processing for sub-THz bands.

\subsubsection{Innovative Network Architectures}

A typical sub-THz communication system will consist of a point-to-point MIMO channel with low rank, due to the directive transmission/reception and lack of scattering in higher bands. The multiple antenna technology can provide three main benefits: beamforming gain, spatial diversity, and spatial multiplexing. However, not all of these benefits can be utilized in every situation. A beamforming gain can be achieved even if the channel has low rank, provided that the CSI is available.

Spatial diversity can be utilized to improve the reliability but requires a MIMO channel with (partially) independent fading between the antennas, which is not the case for low-rank MIMO channels. To combat this deficiency, it will be of interest to consider distributed antenna deployments, for example, based on the cell-free massive MIMO methodology described in Section~\ref{section:cell-free}. Spatial multiplexing also requires a high-rank channel and that can be achieved in the same way. Alternatively, the deployment of many \textcolor{black}{IRSs} (described in Section~\ref{sec:IRS}) can improve the propagation conditions in similar ways.

\textcolor{black}{\subsection{MIMO for Rural Areas}}
Supplying broadband wireless access in rural areas \textcolor{black}{may become} very important for beyond-5G. A cellular massive MIMO BS can provide services for $3,000$ homes in a rural area with similar data rates as in cable- or fiber-based access with the set of parameters given in \cite{Marzetta2016a}. The fading channel in rural \textcolor{black}{areas} is typically LoS, which may reduce the rank of channel matrix. As a result, novel network architecture with MIMO for rural area should be considered. For example, a promising solution could be UAV-based MIMO communication introduced in Section \ref{se:UAV-MIMO}. Furthermore, effective MIMO beamforming techniques \textcolor{black}{can be} utilized in multiple \textcolor{black}{spatially-separated} high altitude platform drones to exploit spatial multiplexing and boost spectral efficiency for ground users in rural areas \cite{mohammed2011role}.

\section{Conclusion}
\label{sec:conclusions}

\textcolor{black}{Multiple antenna technology has become mainstream with 5G, where it plays a key role in
significantly improving capacity, coverage, and QoS over legacy cellular networks. While there
are still many practical aspects that must be dealt with before 5G can reach its full potential, it
is never too early to search for new multiple antenna technologies for beyond-5G applications.
In addition to providing revolutionary performance gains to beyond-5G networks, the new
technologies must also provide orders-of-magnitude gain in energy efficiency at a reasonable
\textcolor{black}{cost to} enable scalable ubiquitous connectivity.}

\textcolor{black}{In this survey, we outline three very promising beyond-5G research directions: Cell-free
massive MIMO, beamspace MIMO, and IRS. Reference points set by recent technical advances
in conjunction with historic perspectives are presented in terms of system models, performance
analyses, signal processing schemes, and deployment visions. Importantly, we provide in-depth
discussions on crucial open problems in each of these areas. From a broader perspective, networks
are becoming increasingly complex and heterogeneous in the future, with conventional massive
MIMO technology being combined with distributed cell-free deployments, supported by IRS,
and signal processing simplifications enabled by the beamspace methodology. These technologies
may be used in both conventional frequency bands and new sub-THz bands, and the networks
might include UAVs for improved coverage. Even after decades of research and development,
we believe that multiple antenna technology will remain a very important and exciting
research avenue for beyond-5G systems.}

\bibliographystyle{IEEEtran}
\bibliography{IEEEabrv,CFmMIMO,refs,DJL_refs,OFDMA-AF,Kwan_other}

\begin{IEEEbiography}[{\includegraphics[width=1in,height=1.25in,clip,keepaspectratio]{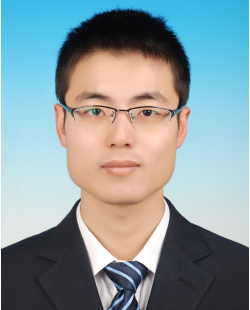}}]
	{Jiayi Zhang}(S'08--M'14) received the B.Sc. and Ph.D. degree of Communication Engineering from Beijing Jiaotong University, China in 2007 and 2014, respectively. Since 2016, he has been a Professor with School of Electronic and Information Engineering, Beijing Jiaotong University, China. From 2014 to 2016, he was a Postdoctoral Research Associate with the Department of Electronic Engineering, Tsinghua University, China. From 2014 to 2015, he was also a Humboldt Research Fellow in Institute for Digital Communications, Friedrich-Alexander-University Erlangen-N\"urnberg (FAU), Germany. From 2012 to 2013, he was a visiting scholor at the Wireless Group, University of Southampton, United Kingdom. His current research interests include massive MIMO, large intelligent surface, communication theory and applied mathematics.
	
	Dr. Zhang received the Best Paper Awards at the WCSP 2017 and IEEE APCC 2017. He was recognized as an exemplary reviewer of the \textsc{IEEE Communications Letters} in 2015 and 2016. He was also recognized as an exemplary reviewer of the \textsc{IEEE Transactions on Communications} in 2017. He was the Lead Guest Editor of the special issue on ``Multiple Antenna Technologies for Beyond 5G" of the \textsc{IEEE Journal on Selected Areas in Communications}. He currently serves as an Associate Editor for \textsc{IEEE Transactions on Communications}, \textsc{IEEE Communications Letters}, \textsc{IEEE Access} and \textsc{IET Communications}.
\end{IEEEbiography}

\begin{IEEEbiography}[{\includegraphics[width=1in,height=1.25in,clip,keepaspectratio]{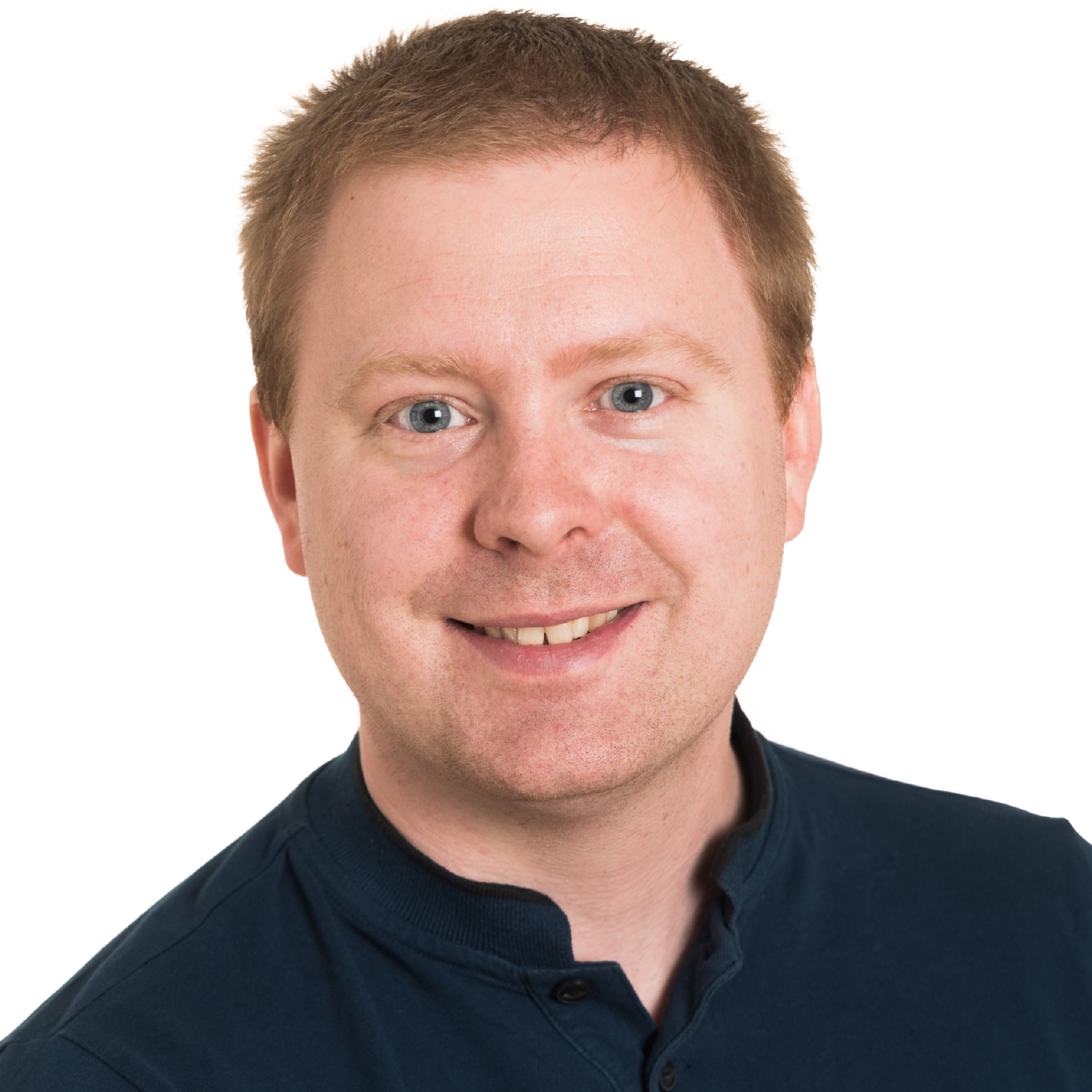}}]
{Emil Bj\"ornson} (S'07-M'12-SM'17) received the M.S. degree in engineering mathematics from Lund University, Sweden, in 2007, and the Ph.D. degree in telecommunications from the KTH Royal Institute of Technology, Sweden, in 2011. From 2012 to 2014, he held a joint post-doctoral position at the Alcatel-Lucent Chair on Flexible Radio, SUPELEC, France, and the KTH Royal Institute of Technology. He joined Link\"oping University, Sweden, in 2014, where he is currently an Associate Professor and a Docent with the Division of Communication Systems.

He has authored the textbooks \emph{Optimal Resource Allocation in Coordinated Multi-Cell Systems} (2013) and \emph{Massive MIMO Networks: Spectral, Energy, and Hardware Efficiency} (2017). He is dedicated to reproducible research and has made a large amount of simulation code publicly available. He performs research on MIMO communications, radio resource allocation, machine learning for communications, and energy efficiency. Since 2017, he has been on the Editorial Board of the \textsc{IEEE Transactions on Communications} and the \textsc{IEEE Transactions on Green Communications and Networking} since 2016.

He has performed MIMO research for over ten years and has filed more than twenty MIMO related patent applications. He has received the 2014 Outstanding Young Researcher Award from IEEE ComSoc EMEA, the 2015 Ingvar Carlsson Award, the 2016 Best Ph.D. Award from EURASIP, the 2018 IEEE Marconi Prize Paper Award in Wireless Communications, the 2019 EURASIP Early Career Award, the 2019 IEEE Communications Society Fred W. Ellersick Prize, and the 2019 IEEE Signal Processing Magazine Best Column Award. He also co-authored papers that received Best Paper Awards at the conferences, including WCSP 2009, the IEEE CAMSAP 2011, the IEEE WCNC 2014, the IEEE ICC 2015, WCSP 2017, and the IEEE SAM 2014.
\end{IEEEbiography}

\begin{IEEEbiography}[{\includegraphics[width=1in,height=1.25in,clip,keepaspectratio]{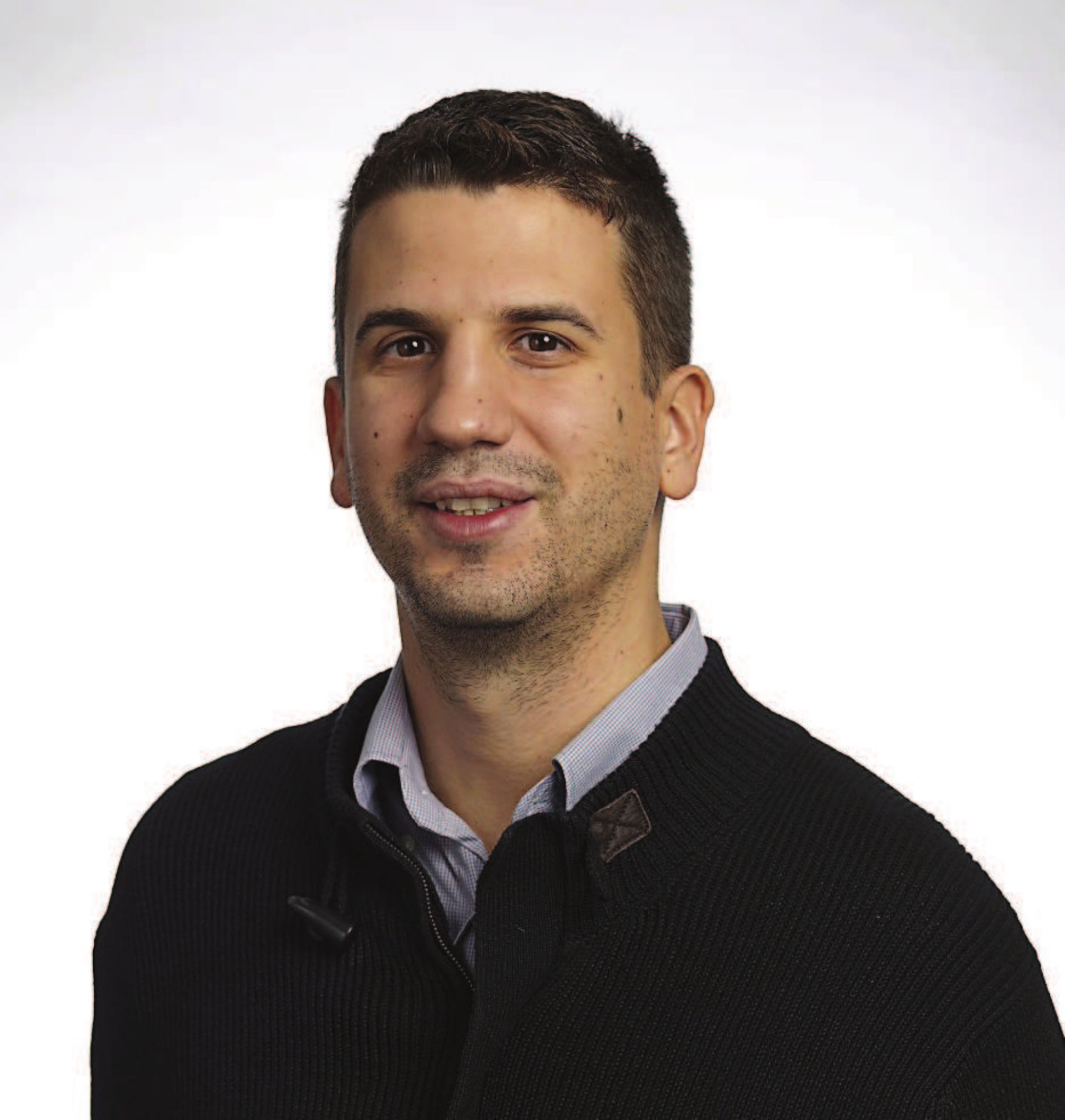}}]
{Michail Matthaiou}(S'05--M'08--SM'13) was born in Thessaloniki, Greece in 1981. He obtained the Diploma degree (5 years) in Electrical and Computer Engineering from the Aristotle University of Thessaloniki, Greece in 2004. He then received the M.Sc. (with distinction) in Communication Systems and Signal Processing from the University of Bristol, U.K. and Ph.D. degrees from the University of Edinburgh, U.K. in 2005 and 2008, respectively. From September 2008 through May 2010, he was with the Institute for Circuit Theory and Signal Processing, Munich University of Technology (TUM), Germany working as a Postdoctoral Research Associate. He is currently a Reader (equivalent to Associate Professor) in Multiple-Antenna Systems at Queen's University Belfast, U.K. after holding an Assistant Professor position at Chalmers University of Technology, Sweden. His research interests span signal processing for wireless communications, massive MIMO systems, hardware-constrained communications, mm-wave systems and deep learning for communications.

Dr. Matthaiou and his coauthors received the IEEE Communications Society (ComSoc) Leonard G. Abraham Prize in 2017. He was awarded the prestigious 2018/2019 Royal Academy of Engineering/The Leverhulme Trust Senior Research Fellowship and recently received the 2019 EURASIP Early Career Award. His team was also the Grand Winner of the 2019 Mobile World Congress Challenge. He was the recipient of the 2011 IEEE ComSoc Best Young Researcher Award for the Europe, Middle East and Africa Region and a co-recipient of the 2006 IEEE Communications Chapter Project Prize for the best M.Sc. dissertation in the area of communications. He has co-authored papers that received best paper awards at the 2018 IEEE WCSP and 2014 IEEE ICC and was an Exemplary Reviewer for \textsc{IEEE Communications Letters} for 2010. In 2014, he received the Research Fund for International Young Scientists from the National Natural Science Foundation of China. He is currently the Editor-in-Chief of Elsevier Physical Communication and a Senior Editor for \textsc{IEEE Wireless Communications Letters}. In the past, he was an Associate Editor for the \textsc{IEEE Transactions on Communications} and Associate Editor/Senior Editor for \textsc{IEEE Communications Letters}.
\end{IEEEbiography}

\begin{IEEEbiography}[{\includegraphics[width=1in,height=1.25in,clip,keepaspectratio]{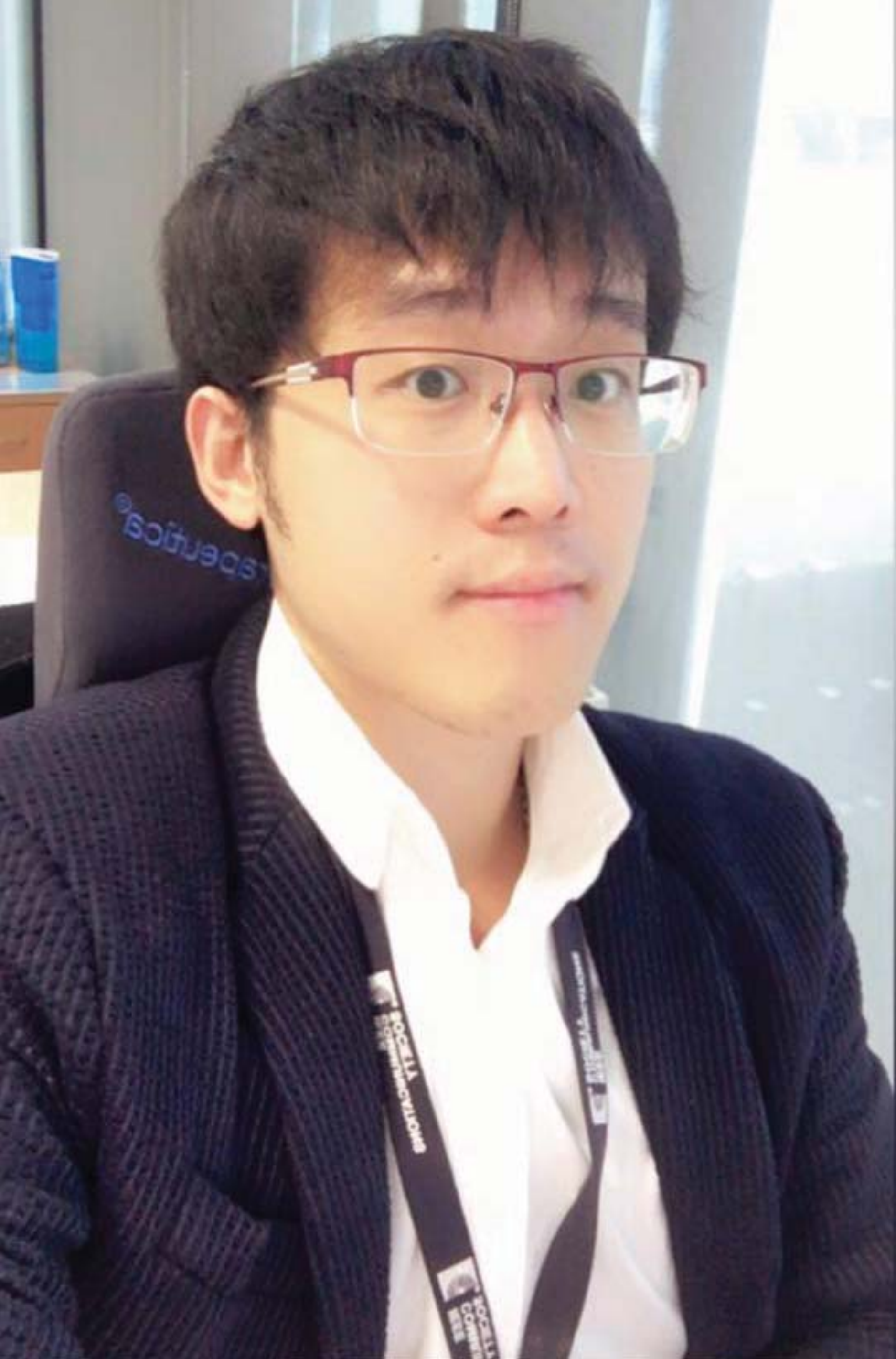}}]
{Derrick Wing Kwan Ng}(S'06-M'12-SM'17)  received the bachelor degree with first-class honors and the Master of Philosophy (M.Phil.) degree in electronic engineering from the Hong Kong University of Science and Technology (HKUST) in 2006 and 2008, respectively. He received his Ph.D. degree from the University of British Columbia (UBC) in 2012. He was a senior postdoctoral fellow at the Institute for Digital Communications, Friedrich-Alexander-University Erlangen-N\"urnberg (FAU), Germany. He is now working as a Senior Lecturer and a Scientia Fellow at the University of New South Wales, Sydney, Australia.  His research interests include convex and non-convex optimization, physical layer security, IRS-assisted communication, UAV-assisted communication, wireless information and power transfer, and green (energy-efficient) wireless communications.

Dr. Ng received the Best Paper Awards at the IEEE TCGCC Best Journal Paper Award 2018, INISCOM 2018, IEEE International Conference on Communications (ICC) 2018,  IEEE International Conference on Computing, Networking and Communications (ICNC) 2016,  IEEE Wireless Communications and Networking Conference (WCNC) 2012, the IEEE Global Telecommunication Conference (Globecom) 2011, and the IEEE Third International Conference on Communications and Networking in China 2008.  He served as an editorial assistant to the Editor-in-Chief of the \textsc{IEEE Transactions on Communications} from Jan. 2012 to Dec. 2019. He is now serving as an editor for the \textsc{IEEE Transactions on Communications},  the \textsc{IEEE Transactions on Wireless Communications}, and an area editor for the \textsc{IEEE Open Journal of the Communications Society}. Also, he was listed as a Highly Cited Researcher by Clarivate Analytics in 2018 and 2019.
\end{IEEEbiography}

\begin{IEEEbiography}[{\includegraphics[width=1in,height=1.25in,clip,keepaspectratio]{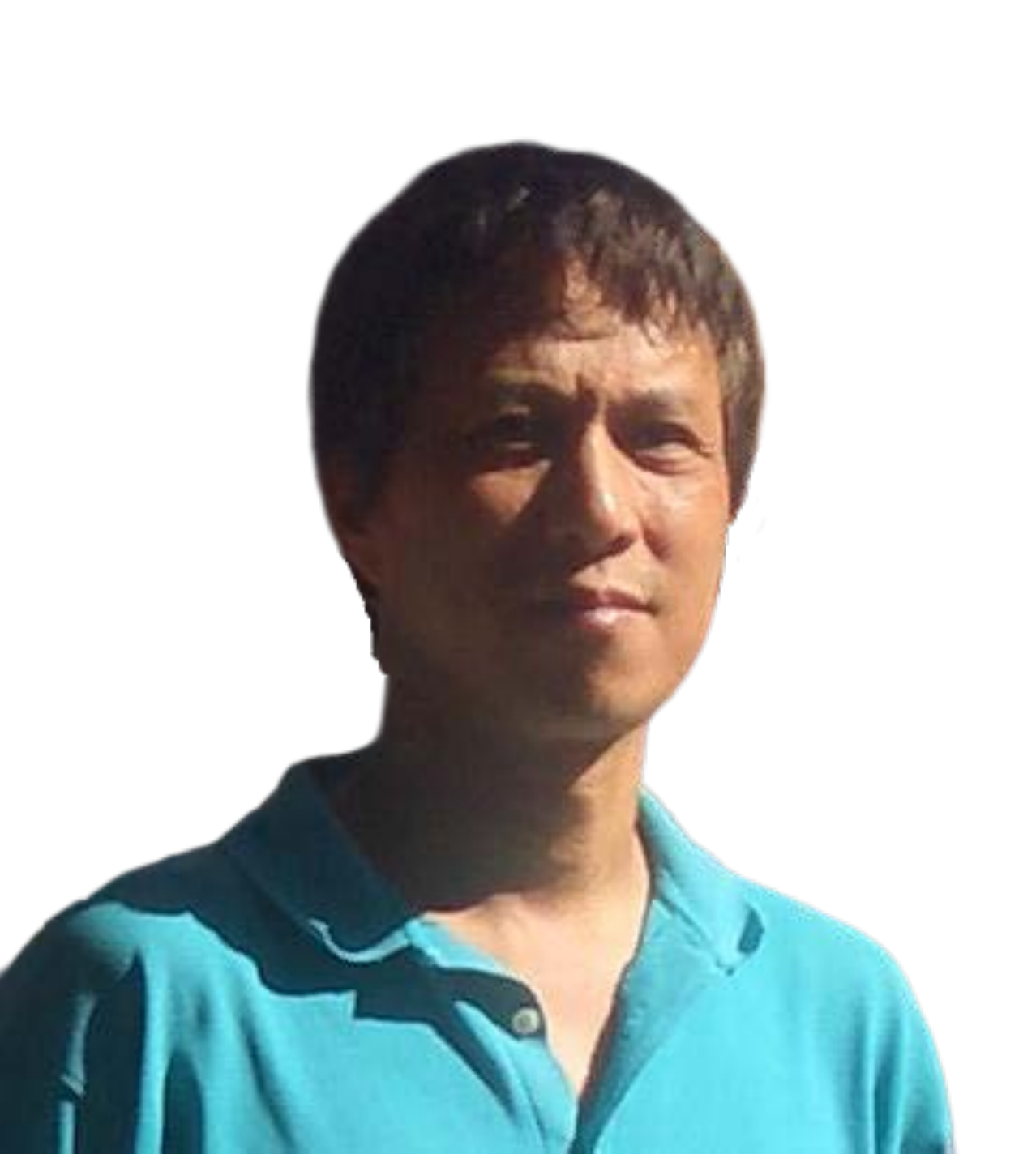}}]
{Hong Yang}(SM'17) received the Ph.D. degree in applied mathematics from Princeton University, Princeton,
NJ, USA. He was involved in academia and a start-up network technology company before joining Lucent Technologies and Alcatel-Lucent, where he was with the Wireless Design Center, the Systems Engineering Department, and Bell Labs Research.
He is currently a member of Technical Staff with the Mathematics of Networks and Communications Research Department, Nokia Bell Labs, Murray Hill, NJ, USA, where he conducts research in communications networks. He has co-authored many research papers in wireless communications, applied mathematics, control theory, and financial economics. He co-invented many U.S. and international patents. He co-authored the book \emph{Fundamentals of Massive MIMO} (Cambridge University Press, 2016).
\end{IEEEbiography}

\begin{IEEEbiography}[{\includegraphics[width=1in,height=1.25in,clip,keepaspectratio]{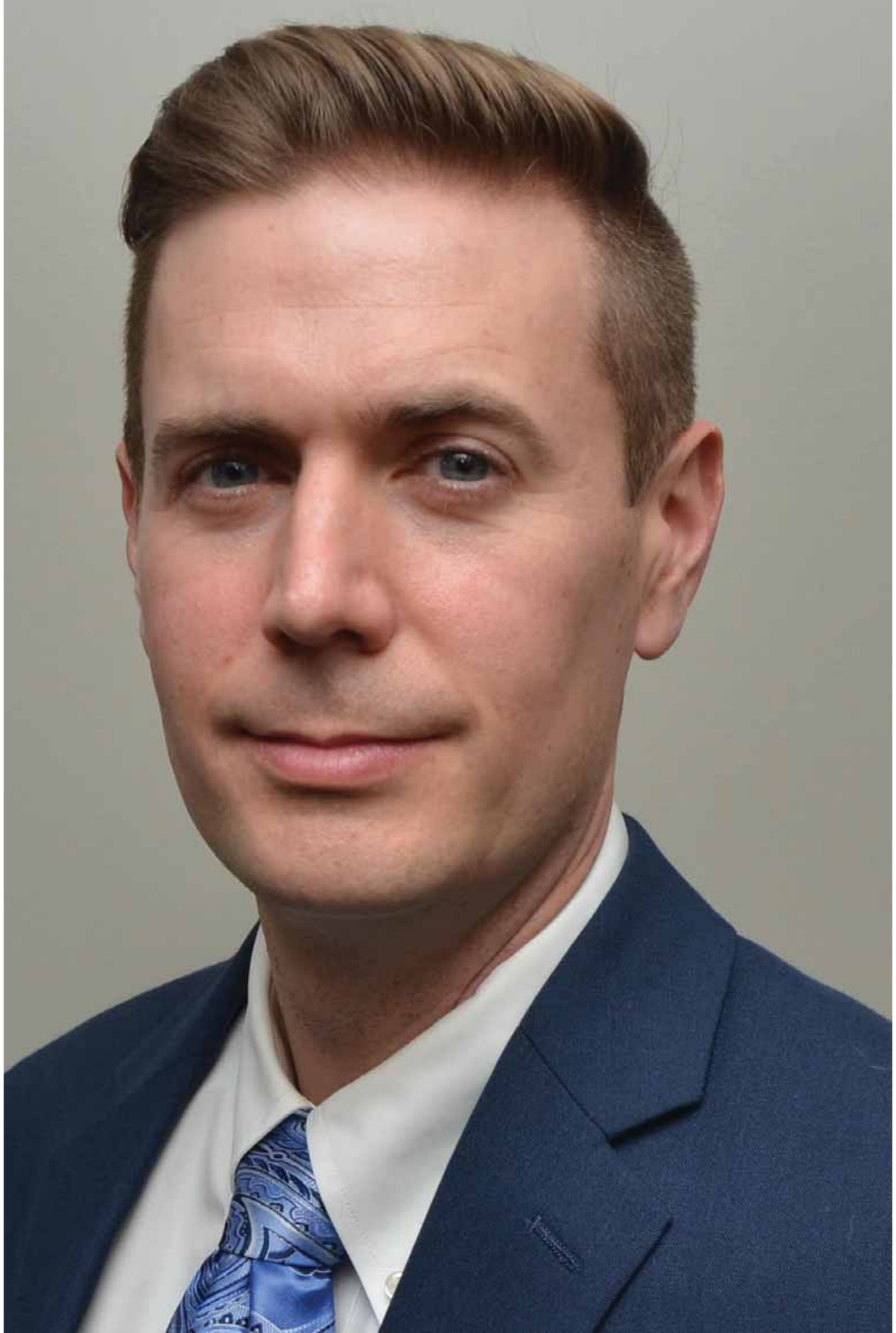}}]
{David J. Love} (S'98--M'05--SM'09--F'15)  received the B.S. (with highest honors), M.S.E., and Ph.D. degrees in electrical engineering from the University of Texas at Austin in 2000, 2002, and 2004, respectively.  Since 2004, he has been with the School of Electrical and Computer Engineering at Purdue University, where he is now the Nick Trbovich Professor of Electrical and Computer Engineering and leads the College of Engineering Preeminent Team on Efficient Spectrum Usage.  He currently serves as a Senior Editor for \textsc{IEEE Signal Processing Magazine} and previously served as an Editor for the \textsc{IEEE Transactions on Communications}, an Associate Editor for the \textsc{IEEE Transactions on Signal Processing}, and a guest editor for special issues of the \textsc{IEEE Journal on Selected Areas in Communications} and the \textsc{EURASIP Journal on Wireless Communications and Networking}. He is a member of the Executive Committee for the National Spectrum Consortium.  He holds 31 issued U.S. patent filings.  His research interests are in the design and analysis of broadband wireless communication systems, 5G wireless systems, multiple-input multiple-output (MIMO) communications, millimeter wave wireless, software defined radios and wireless networks, coding theory, and MIMO array processing.

Dr. Love was named a Thomson Reuters Highly Cited Researcher (2014 and 2015), is a Fellow of the Royal Statistical Society, and has been inducted into Tau Beta Pi and Eta Kappa Nu.  Along with his co-authors, he won best paper awards from the IEEE Communications Society (2016 IEEE Communications Society Stephen O. Rice Prize), the IEEE Signal Processing Society (2015 IEEE Signal Processing Society Best Paper Award), and the IEEE Vehicular Technology Society (2009 IEEE Transactions on Vehicular Technology Jack Neubauer Memorial Award).
\end{IEEEbiography}

\end{document}